\documentclass[prl,showpacs,twocolumn]{revtex4}
\usepackage{color,epsfig}
\usepackage{bm}
\usepackage{amsmath}

\begin{document}

\title{Quasispecies theory for finite populations}

\author{Jeong-Man Park$^{1,2}$, Enrique Mu\~noz$^1$, and
Michael W.\ Deem$^1$
}

\affiliation{
\hbox{}$^1$Department of Physics \& Astronomy,
Rice University, Houston,Texas 77005--1892, USA\\
\hbox{}$^2$Department of Physics, The Catholic University of
Korea, Bucheon 420-743, Korea}

\begin{abstract}
We present stochastic, finite-population formulations
of the Crow-Kimura and Eigen models of quasispecies theory,
for fitness functions that depend in an arbitrary way on
the number of mutations from the wild type.
We include back mutations in our description. 
We show that the fluctuation of the population numbers about 
the average values are exceedingly large in these physical
models of evolution.  We further show that horizontal gene transfer
reduces by orders of magnitude the fluctuations in the population numbers
and reduces the accumulation of deleterious mutations 
in the finite population due to Muller's ratchet. 
Indeed the population sizes
needed to converge to the infinite population limit are often larger
than those found in nature for smooth fitness functions in the
absence of horizontal gene transfer.
These analytical results are derived for the steady-state
by means of a field-theoretic
representation. Numerical results are presented that indicate horizontal
gene transfer speeds up the dynamics of evolution as well.
\end{abstract}


\maketitle

\section{Introduction}

Biological populations in nature are finite. In particular, it
is clear that the
number of individuals in a population
is much smaller than the number of possible genetic sequences, even for
genomes of modest length.
For example, the largest populations observed in biological
systems, RNA viruses, are on the order of $N = 10^{12}$
viral particles within a single infected organism \cite{Moya00}.
These viruses possess a relatively short
genome of length $L \sim 10^{3} - 10^{4}$ bases \cite{Moya00},
and hence the theoretical size of the sequence space is
$4^{L} \sim 10^{6000} \gg N$. Even
the region of phase space for which fitness is high
is typically much larger than the biological population size. 
From this example, it is
clear that no real biological population will be able
to sample the entire sequence space during
evolutionary dynamics \cite{Zhang97},
and therefore finite population
size effects may be important for a realistic description
of evolution \cite{Alves98}. Finite populations with asexual
reproduction are subject to the
``Muller's ratchet'' effect \cite{Muller64}, which is the
tendency to accumulate deleterious mutations in finite populations
\cite{Muller64,Arjan07,Kondrashov88}.
It has been suggested that horizontal gene transfer and recombination
may provide a way to escape Muller's ratchet in small
populations \cite{Chao92,Chao97,Rice01,Adami06},
and this mechanism has been proposed as one
of the evolutionary advantages of sex, despite the additional mutational
load for fitness functions with
positive epistasis
\cite{Muller64,Otto02,Arjan07,Kondrashov88,Kondrashov82,Kouyos06,Kouyos07,Rice01}. The role of the finite population size in the Muller's ratchet effect
has been previously studied by the traveling-wave approximation
\cite{Rouzine03,Rouzine08}. This 
theoretical approach introduces an approximate treatment, 
by assuming deterministic dynamics for the bulk of
the population, but stochastic dynamics for the edge composed of the
class of highest fitness genotypes. The deterministic component of this theory,
which considers single point mutations coupled to replication,
is similar to 
traditional quasispecies models for infinite populations. These
previous studies considered only linear fitness functions
and analyzed in detail the case of no back mutations, an approximation
which changes the dynamics and leads to a different
steady-state distribution.
An exception is the model in Ref.\ \cite{Cohen05}, which
presents a mean-field approximation
which incorporates single back mutations in a linear fitness.

We here include back mutations and consider fitness functions that
depend in an arbitrary way on the number of mutations from the
wild type in our exact description.  

Quasispecies models for molecular evolution, represented
by the Crow-Kimura model \cite{Kimura70} and the
Eigen model \cite{Eigen71,Eigen88,Eigen89,Biebricher05},
are traditionally formulated
in the language of chemical kinetics. That is, they describe
the basic processes of mutation and selection
in an infinite population of self-replicating, information
encoding molecules such as RNA or DNA, which
are assumed to be drawn from a binary alphabet (e.g. purines/pyrimidines).
These models exhibit a phase transition in the infinite
genome limit \cite{Eigen71,Eigen88,Eigen89,Tarazona92,Leuthausser87,Biebricher05,Franz97,Park06,Saakian06}, separating an organized or quasispecies
phase from a disordered phase. This phase transition occurs when the
mutation rate exceeds a critical value, which depends on the nature
of the fitness function \cite{Park06,Munoz08}. The phase transition is
usually of first order for binary alphabets \cite{Park06,Munoz08}, but
it is of higher order for smooth fitness
functions in larger alphabets \cite{Munoz09}. 
The quasispecies is composed by
a collection of nearly neutral mutants, that is, a cloud
of closely related individuals sharing similar fitness values, 
rather than by a single
sequence type. Despite its abstract character, the quasispecies
model has been successfully applied to interpret experimental
studies in RNA viruses \cite{Domingo78,Domingo05,Ortin80,Domingo85}.

\section{Finite population effects in the Crow-Kimura Model}

\begin{figure}[tbp]
\centering
\epsfig{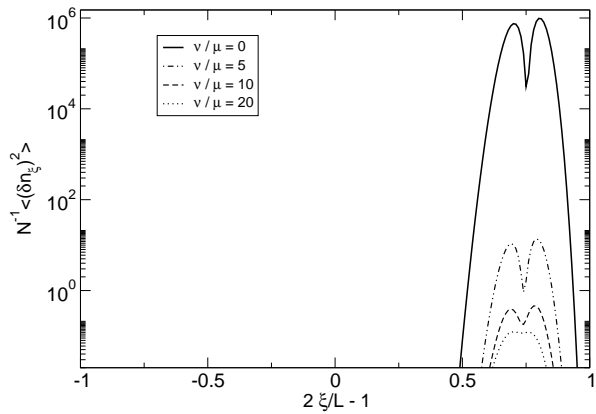}
\caption{Fluctuation in the number of individuals with a given
sequence composition.  The quadratic fitness
is used in the parallel model, with $L=200$ and $k = 4.0$.
The theory is obtained from Eqs.\ (\ref{eq11}) and (\ref{eq12}). 
Fluctuations decrease by orders of magnitude with increasing
horizontal gene transfer rate, $\nu$.
\label{fig1}
}
\end{figure}

\begin{figure}[tbp]
\centering
\epsfig{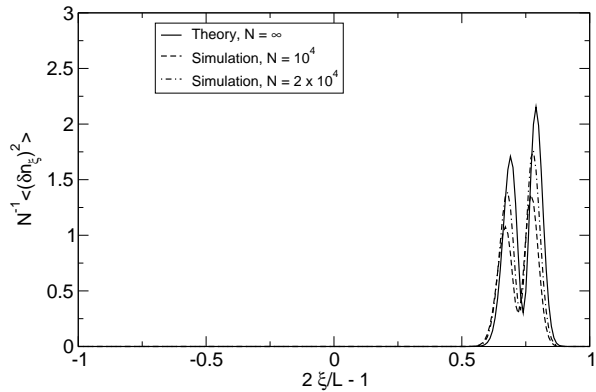}
\caption{
Stochastic results
obtained by averaging over 50 independent Gillespie simulations, are
shown and compared with analytical
theory, for $\nu=7.0$. 
\label{fig2}
}
\end{figure}

In the infinite population limit,
the mean field approach that is customary in chemical
kinetics is justified, and the evolution of the probability
distribution of sequence types can be described by a deterministic
system of differential equations. This mean field
approach cannot capture the fluctuations in the numbers
of individuals with different sequences,
which are a consequence of the stochastic dynamics
of the process. An accurate description of all aspects of
a finite population
therefore requires a master equation formulation \cite{Alves98}.
We here consider arbitrary fitness functions. The special case
of linear fitness functions $f(\xi) = a\xi$, have been analyzed 
in \cite{Rouzine03,Rouzine08,Cohen05}.

We consider a finite population, composed of $N<\infty$
binary purine/pyrimidine
sequences, of length $L$. 
The terms in the master equation for the Crow-Kimura, or parallel,
model are i) a replication term, whereby each
individual of sequence $S_i$ reproduces at a rate $L f(S_i)$ and
the offspring replaces a random member of the population,
ii) a mutation term, whereby each base in a sequence mutates
at a rate $\mu$ per unit time,
and iii) a horizontal gene transfer term, whereby bases
in a sequence are replaced at rate $\nu$ per unit time
with bases randomly chosen from the population.
We assume that the replication
rate, or microscopic fitness, is a
function of the Hamming
distance from the wild-type genome, and hence of the
one-dimensional coordinate $0\le \xi \le L$ representing
the alignment of an individual's sequence with the wild type.
The master equation can be exactly projected onto the $\xi$ coordinate
and defines the rates at which the sequences of individuals
change with time due to replication, mutation, and
horizontal gene transfer. We define $(1+u)/2$ to be the
probability of a wild type letter in the sequence, 
$\rho_{\pm}=(1\pm u)/2$ 
is the probability of inserting a wild-type or non-wild-type
letter by horizontal gene transfer \cite{Park07,Munoz08},
and 
\begin{eqnarray}
u = \frac{1}{N}\sum_{\xi=0}^{L}\left( 2\xi/L - 1\right)n_{\xi}
\label{eq1}
\end{eqnarray}
is the `average base composition,' where $n_{\xi}$ is the
number of individuals at coordinate $\xi$.

\begin{figure}[tbp]
\centering
\epsfig{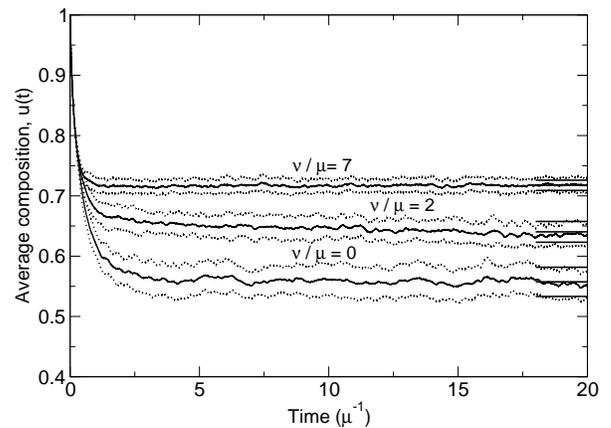}
\caption{
The average composition as a function of time, averaged over 50 independent
Gillespie simulations, with population size $N = 10^4$ (solid curves).
Also shown are one standard deviation envelopes $\pm\sigma(t)$ (dotted
curves).
The steady-state averages $\langle u \rangle \pm \sqrt{\langle(\delta u)^{2}
\rangle}$ are displayed as solid lines for reference.
\label{fig3}
}
\end{figure}

We formulate the master equation for the probability distribution 
$P(\{n_{\xi}\};t)$, 
as a function of the set of occupation numbers $\{n_{\xi}\}_{0\leq\xi\leq L}$.
As in the classical, infinite population Crow-Kimura model \cite{Kimura70}, 
we consider point mutation with rate $\mu$, 
and replication with a rate $r(\xi)=Lf(\xi)$, 
while preserving the population size $N$. In addition, we consider
horizontal gene transfer of single letters between an individual sequence and
the population, with rate $\nu$.

The master equation describing
this process is
\begin{eqnarray}
\frac{\partial}{\partial t}P(\{n_{\xi}\})
&=&
\frac{1}{N}\sum_{\xi\ne\xi'}
r(\xi)[(n_{\xi}-1)(n_{\xi'}+1)\nonumber\\
&\times & P(n_{\xi}-1,n_{\xi'}+1)
- n_{\xi}n_{\xi'}
P(\{n_{\xi}\})]\nonumber\\
&+& \mu\sum_{\xi=0}^{L}[(L-\xi)(n_{\xi}+1)P(n_{\xi}+1,n_{\xi+1}-1)\nonumber\\
&+& \xi(n_{\xi}+1)P(n_{\xi-1}-1,n_{\xi}+1)-L n_{\xi}\nonumber\\
&\times & P(\{n_{\xi}\}) ]
+\nu\sum_{\xi=0}^{L}[\rho_{+}(L-\xi)(n_{\xi}+1)\nonumber\\
&\times &P(n_{\xi}+1,n_{\xi+1}-1)
+ \xi\rho_{-}(n_{\xi}+1)\nonumber\\
&\times & P(n_{\xi-1}-1,n_{\xi}+1)-
n_{\xi}\{\rho_{+}(L-\xi)\nonumber\\
&+& \rho_{-}\xi\}P(\{n_{\xi}\})]
\label{eq2}
\end{eqnarray}

Note that this exact master equation includes 'back mutations'
often ignored in the literature \cite{Rouzine03,Rouzine08}. Note
that the approximation of setting back mutations to zero leads
to both different dynamics and a different steady-state.

\subsection{Mapping to a field theory}

We seek analytical expressions for 
the fluctuations in number of individuals with given sequence compositions in
the finite population parallel model.
We derive these results by
means of a field-theoretic method \cite{Park06,Peliti85,Mattis98}.
This approach provides
a system of coupled differential equations
for the probability distribution and the fluctuation
of numbers of individuals with given sequence composition, whose
computational solution is essentially instantaneous.
These results 
give us the fluctuation and correlation in
population numbers and 
are an exact expansion in the inverse of the
population size.  
\begin{figure}[tbp]
\centering
\epsfig{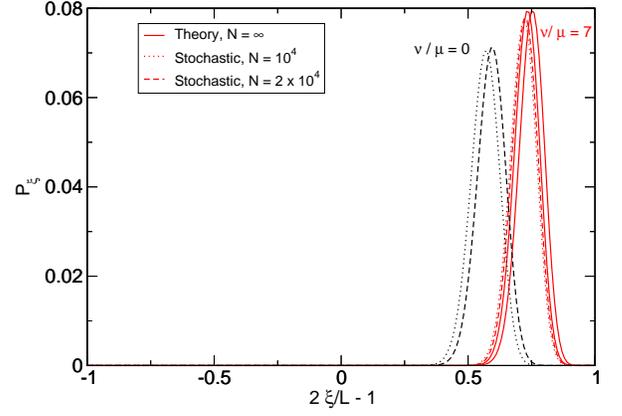}
\caption{(Color online) Finite population versus infinite population results
for the probability distribution of the
parallel model with quadratic fitness.
Note that the Muller's ratchet phenomenon, whereby fitness is reduced for
finite populations, is greatly suppressed for $\nu>0$.
Here $k=4$ and $L = 200$, and
the stochastic results are obtained
by averaging over 50 independent numerical experiments.
\label{fig4}
}
\end{figure}
We introduce an exact representation of the classical master equation
in terms of a many-body quantum theory \cite{Park06}. 
For that purpose,
we define the population state vector
\begin{eqnarray}
|\Psi(t)\rangle = \sum_{\{n_{\xi}\}}P(\{n_{\xi}\};t)|\{n_{\xi}\}\rangle
\label{eq3}
\end{eqnarray}
with 
\begin{eqnarray}
|\{n_{\xi}\}\rangle = |n_{0},n_{1},\ldots,n_{L}\rangle = 
\prod_{\xi=0}^{L}\otimes |n_{\xi}\rangle
\label{eq4} 
\end{eqnarray}
This population
state vector evolves according to a Schr\"odinger equation
in imaginary time, 
\begin{eqnarray}
\frac{d}{dt}|\Psi(t)\rangle = -\hat{H}|\Psi(t)\rangle
\label{eq5}
\end{eqnarray}
which possesses the formal solution
\begin{eqnarray}
|\Psi(t)\rangle = e^{-\hat{H}t}|\Psi(0)\rangle
\label{eq6}
\end{eqnarray}
with $|\Psi(0)\rangle = |\{n_{\xi}^{0}\}\rangle$ 
representing the initial configuration of the population. 
The master equation is written in second quantized form, with a
Hamiltonian expressed in terms
of boson creation and destruction operators
$[\hat{a}_{\xi},\hat{a}_{\xi^{'}}^{\dagger}]=\delta_{\xi,\xi^{'}}$, whose
action over the occupation number vectors is defined by 
$\hat{a}_{\xi}|n_{\xi}\rangle = n_{\xi}|n_{\xi}-1\rangle$, and 
$\hat{a}_{\xi}^{\dagger}|n_{\xi}\rangle=|n_{\xi}+1\rangle$. The Hamiltonian
is given by
\begin{eqnarray}
-\hat{H} 
&=& 
\frac{1}{N}\sum_{\xi,\xi^{'}=0}^{L} L f(\xi)\hat{a}_{\xi}^{\dagger}(
\hat{a}_{\xi}^{\dagger}-\hat{a}_{\xi^{'}}^{\dagger})
\hat{a}_{\xi}\hat{a}_{\xi^{'}}
\nonumber \\ &+& 
\mu\sum_{\xi=0}^{L}[(L-\xi)(\hat{a}_{\xi+1}^{\dagger}
-\hat{a}_{\xi}^{\dagger})\hat{a}_{\xi}+\xi(\hat{a}_{\xi-1}^{\dagger}-
\hat{a}_{\xi}^{\dagger})\hat{a}_{\xi}]
\nonumber\\ &+&
\nu\sum_{\xi=0}^{L}[\rho_{+}(L-\xi)(\hat{a}_{\xi+1}^{\dagger}-
\hat{a}_{\xi}^{\dagger})\hat{a}_{\xi}+
\rho_{-}\xi(\hat{a}_{\xi-1}^{\dagger}\nonumber\\
&-&\hat{a}_{\xi}^{\dagger})\hat{a}_{\xi}]
\label{eq7}
\end{eqnarray}
The terms proportional to 
$f$ represent replication,
$\mu$ represent mutation, and
$\nu$ represent horizontal gene transfer.
The
population average of a (normal-ordered) classical observable, represented by
the operator $F(\{\hat{a}_{\xi}\})$, is obtained by the inner product
with the "sum" \cite{Peliti85} bra $\langle \cdot | = \langle 0 |
\left(\prod_{\xi=0}^{L}e^{\hat{a}_{\xi}}\right)$,
\begin{eqnarray}
\langle F \rangle = \langle \cdot|F(\{\hat{a}_{\xi} \})|\Psi(t) \rangle
=\langle \cdot| F(\{\hat{a}_{\xi}\})e^{-\hat{H}t}|\{n_{\xi}^{0}\}\rangle
\label{eq8}
\end{eqnarray}
A  Trotter factorization is introduced for the evolution
operator $e^{-\hat{H}t}$ in a basis of coherent states, defined
as $\hat{a}_{\xi}|z_{\xi}\rangle = z_{\xi}|z_{\xi}\rangle$. 
This procedure
leads to a path integral representation \cite{Park06,Munoz08,Munoz09},
\begin{eqnarray}
\langle F \rangle = 
\int [{\mathcal{D}}z^{*}{\mathcal{D}}z]F
(\{z_{\xi}
(t/\epsilon)
\})
e^{-S[\{z^{*}\},\{z\}]}.
\label{eq9}
\end{eqnarray}
Here $z$ are the coherent state field of the second quantized theory
of the parallel model, and $S$ is the corresponding action.
The action in the exponent of Eq.\ (\ref{eq9}) is given, after
the change of variables $z^{*} = 1 + \bar{z}$, in continuous time by
\begin{widetext}
\begin{eqnarray}
S[\{\bar{z}\},\{z\}]&=& \sum_{\xi=0}^{L}\int_{0}^{T}dt^{'}
\bigg\{\bigg[\bar{z}_{\xi}(t^{'})z_{\xi}(t^{'})
-n_{\xi}^{0}\ln[1+\bar{z}_{\xi}(t^{'})]\bigg]\delta(t^{'})
+\bar{z}_{\xi}\frac{\partial z_{\xi}}{\partial t^{'}}
-\mu[(L-\xi)\bar{z}_{\xi+1}
+\xi\bar{z}_{\xi-1}-L\bar{z}_{\xi}]z_{\xi}\nonumber\\
&&-\nu[(L-\xi)\rho_{+}
\bar{z}_{\xi+1}+\xi\rho_{-}\bar{z}_{\xi-1}-
\{(L-\xi)\rho_{+}+\xi\rho_{-}\}\bar{z}_{\xi}]z_{\xi}
-\frac{1}{N}\sum_{\xi^{'}=0}^{L}
L f(\xi)(1+\bar{z}_{\xi})[\bar{z}_{\xi}-\bar{z}_{\xi^{'}}]z_{\xi}
z_{\xi^{'}}\bigg\}
\label{eq10}
\end{eqnarray}
\end{widetext}
In the limit of a large population,
we look for a saddle-point in the action Eq.\ (\ref{eq10}). From the
condition $\left.\frac{\delta S}{\delta z_{\xi}(t)}\right|_{c}=0$, we
obtain $\bar{z}_{\xi}^{c}(t)=0$.
From the condition 
$\left.\frac{\delta S}{\delta \bar{z}_{\xi}(t)}\right|_{c}=0$, we find the 
saddle-point solution 
$z_{\xi}^{c}(t) = N P_{\xi}(t)$, where $P_{\xi}$  
satisfies the  differential equation 
for infinite population 
quasispecies theory, generalized to include
horizontal gene transfer \cite{Park07,Munoz08}:
\begin{eqnarray}
\frac{d}{dt}P_{\xi} &=& \mu[(L-\xi+1)P_{\xi-1}+(\xi+1)P_{\xi+1}-L P_{\xi}]
\nonumber\\
&+&\nu[\rho_{+}(L-\xi+1)P_{\xi-1}+\rho_{-}(\xi+1)P_{\xi+1}-
\{(L\nonumber\\
&-&\xi)\rho_{+}+\xi\rho_{-} \}P_{\xi}]
+ [r(\xi) - \sum_{\xi'=0}^{L}r(\xi')P_{\xi'}]P_{\xi}
\label{eq11}
\end{eqnarray}
Details are given in Appendix 1.

\subsection{Fluctuations}
To calculate fluctuations, we expand the action up to second order, 
to obtain the correlation matrix 
$\langle\delta z_{\xi}(t)\delta z_{\xi^{'}}(t)\rangle = C_{\xi,\xi^{'}}(t)$,
which 
in continuous time evolves according to 
the Lyapunov equation
\begin{eqnarray}
\frac{d}{dt}C = A C + C A^{T} + B
\label{eq12}
\end{eqnarray}
subject to the initial condition 
$C_{\xi,\xi^{'}}(0)=-n_{\xi}^{0}\delta_{\xi,\xi^{'}}$.
Here, the matrices $A$ and $B$ are defined by 
\begin{eqnarray}
&&[A]_{\xi,\xi^{'}}= \delta_{\xi-1,\xi^{'}}(L-\xi+1)[\mu+\nu\rho_{+}]
+\delta_{\xi,\xi^{'}}[L f(\xi)
\nonumber\\
&&
-
\sum_{\xi_{1}}L f(\xi_{1})P_{\xi_{1}}
-\nu\{(L-\xi)\rho_{+}+\xi\rho_{-}\}
-L\mu]
\nonumber\\
&&
+L [f(\xi)-f(\xi^{'})]P_{\xi}+
\delta_{\xi+1,\xi^{'}}(\xi+1)[\mu+\nu\rho_{-}]
\label{eq13}
\end{eqnarray}
\begin{eqnarray}
[B]_{\xi,\xi^{'}}=\delta_{\xi,\xi^{'}} 2 L f(\xi) N P_{\xi}
-L [f(\xi)+f(\xi^{'})]N P_{\xi} P_{\xi^{'}}
\label{eq14}
\end{eqnarray}
See Appendix 2 for details in the derivation.

The fluctuations in the number of individuals with a given
sequence composition are obtained from the relation
\begin{eqnarray}
\frac{(\delta n_{\xi})^{2}}{N^{2}}=\frac{1}{N}(P_{\xi}
+\frac{1}{N}C_{\xi,\xi})
\label{eq15}
\end{eqnarray}

\subsection{Continuous and discontinuous fitness functions}
We consider two example fitness functions, which exhibit a quasi-species phase
transition in the infinite genome length limit $L\rightarrow\infty$.  
The sharp peak represents the extreme case of the wild type sequence 
replicating at a high rate,
and all other sequences replicating at a single lower rate.
The sharp peak fitness function represents a very strong selective advantage
for the wild type.
For the sharp peak $f(\xi) = A\delta_{\xi,L}$, from Eq. (\ref{eq11}) 
and large $L$, 
we find that the wild-type probability
\begin{eqnarray}
\frac{d}{dt}P_{L} 
\simeq L A P_{L}(1 - P_{L}) - L(\mu + \nu\rho_{-})P_{L}
\label{eq16}
\end{eqnarray}
At steady-state, taking into account that $u = 1 - O(L^{-1})$ for the sharp
peak, we have $\rho_{-} = (1-u)/2 = O(L^{-1})$, and after Eq. (\ref{eq16}) 
we find
\begin{eqnarray}
P_{\xi=L} = \left\{\begin{array}{cc}0, & \frac{\mu}{A}>1\\
1 - \mu/A + O(L^{-1}), & \frac{\mu}{A}<1\end{array}\right.
\label{eq17}
\end{eqnarray}
Notice that the steady-state distribution is not affected by horizontal
gene transfer ($\nu>0$). To obtain the fluctuations in the
probability distribution, we consider Eq.\ (\ref{eq12}) for the
matrix element $C_{L,L}$. The terms $C_{L,L\pm 1}$ are $O(L^{-1})$. We also
notice that $\sum_{\xi_{1}=0}^{L}C_{\xi_{1},L}= -N P_{L}$,
to find that the stationary solution of Eq.\ (\ref{eq12})
is given by
\begin{eqnarray}
0 &=& L A N P_{L}(1 - P_{L}) - \mu L C_{L,L} - \nu\rho_{-} L C_{L,L}\nonumber\\
&+& L A (1 - P_{L}) C_{L,L} - L A N P_{L}^{2} - L A P_{L} C_{L,L}\nonumber\\
&=& A N P_{L}(1 - 2P_{L}) + [(A - \mu - \nu\rho_{-}) - 2 A P_{L}]C_{L,L}
\nonumber\\
\label{eq18}
\end{eqnarray}
From Eq.\ (\ref{eq18}), we have $A - \mu - \nu\rho_{-} = A P_{L}$,
and substituting into Eq.\ (\ref{eq18}) we obtain
\begin{eqnarray}
C_{L,L} = N(1 - 2P_{L})
\label{eq19}
\end{eqnarray}
Substitution of this result into Eq.\ (\ref{eq15}) shows that the 
fluctuation is given by
\begin{eqnarray} 
\langle (\delta n_{\xi=L})^{2}\rangle/N^2 = \left\{\begin{array}{cc}0,&
\mu/A > 1\\ \mu/(NA), & \mu/A < 1\end{array}\right.
\label{eq20}
\end{eqnarray}
a result first given in Ref. \cite{Saakian07} by a different method.

The second fitness function we consider is one for which the
replication rate decreases continuously as a function of the
Hamming distance from the wild type.
In particular, we
choose a quadratic fitness 
$f(\xi) = (k/2) (2 \xi/L-1)^{2}$.
The quadratic fitness represents any continuous fitness
function, for which mutants reproduce more slowly than the wild type,
in a way that depends continuously on the Hamming distance from the
wild type. 
Figure\ \ref{fig1} shows that horizontal gene transfer
reduces by orders of magnitude
the fluctuations in number of individuals with a given
sequence composition, $n_\xi$.
Indeed, a small rate of horizontal
gene transfer is enough to reduce by several orders of magnitude
these fluctuations, as compared to the case without horizontal
gene transfer, $\nu = 0$.

The linear fitness function $f(\xi) = A \xi/L$ was considered in
\cite{Cohen05} and in \cite{Rouzine03,Rouzine08} in the absence of
back mutations. 
The steady-state exhibits no phase transition for the
linear fitness. We skip this example in favor of the forms considered above. 
 
\subsection{Stochastic simulations}
We performed Lebowitz/Gillespie simulations
\cite{Gillespie76,Lebowitz75} in which we explicitly simulate
a population of size $N$ undergoing the stochastic processes
of mutation, horizontal gene transfer, and replication. 
In Fig.\ \ref{fig2} and Fig.\ \ref{fig3}, 
we compare our theory with stochastic
simulations, at different rates of horizontal gene transfer. The
results obtained from stochastic simulations converge toward
the theoretical value calculated from Eqs.\ (\ref{eq11}) and (\ref{eq12})
as the size of the population, $N$, increases. 
Non-zero horizontal gene transfer rates both reduce fluctuations and 
accelerate convergence
towards the infinite-population value of the mean fitness.

In Fig.\ \ref{fig4}, the steady-state
probability distribution obtained from the numerical solution of
Eq.\ (\ref{eq11}) is compared with the distributions obtained
from stochastic simulations, for different sizes, $N$, of the population. 
The convergence with $N$ toward the infinite-population limit is
more rapid for non-zero $\nu$. Indeed for smooth fitness functions, the
infinite population limit is only reached for population sizes larger
than those commonly found in nature. For the discontinuous sharp peak
fitness function, on the other hand, fluctuations are small, 
Eq.\ (\ref{eq20}), and the convergence to the infinite population limit
is rapid.

Another point from Fig.\ \ref{fig3} is that horizontal gene transfer speeds 
up the rate of evolution. We see that the convergence to the steady state
is more rapid for increased horizontal gene transfer rates. Numerical
experiments have shown that the effect of horizontal gene transfer
on the rate of evolution is especially dramatic for rugged fitness
landscapes \cite{Bogarad99,Earl04}. At the local scale, biological
fitness landscapes may be relatively smooth. At larger genetic distances,
however, we expect biological fitness landscapes to be rugged.
Correlations exists in the rugged landscape, and horizontal gene transfer
couples to those correlations in a way that allows evolution to
speed up dramatically \cite{Sun07}. We expect that this speedup of
evolution on rugged landscaped is one of the most significant effects
of horizontal gene transfer in biology.

\begin{figure}[tbp]
\centering
\epsfig{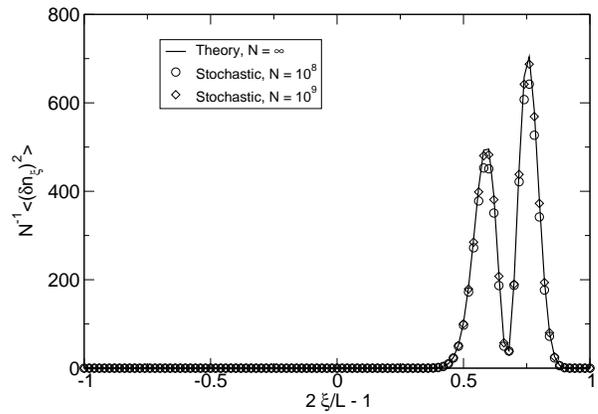}
\caption{Fluctuations in the probability distribution for the Crow-Kimura model,
obtained from stochastic simulations using the Gillespie method (dots and
diamonds) at different sizes of the population, in
the absence of horizontal gene transfer $\nu = 0$. 
Convergence towards the theoretical curve Eq.\ (\ref{eq12}) (solid line)
is observed. Here $L=200$, and the quadratic fitness with $k = 4.0$ and $\mu=1$
was considered.
\label{fig8}
}
\end{figure}

Note that when $\nu=0$ the number fluctuations for the case of
fitness functions for which the population
is not exponentially localized at $\xi = L$ (i.e. continuous fitness
functions) are large in comparison to the fluctuations for a localized 
population, e.g. sharp peak. 
Another way to see this effect is shown in Fig.\ \ref{fig8},
where for $\nu=0$, the convergence to $N\rightarrow\infty$ is slow.

\begin{figure}[tbp]
\centering
\epsfig{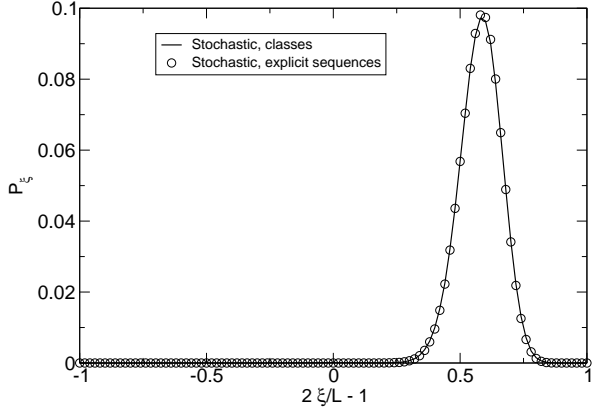}
\caption{Probability distributions for the Crow-Kimura model,
obtained from stochastic simulations using the Gillespie method with
explicit sequences or alternatively with Hamming distance classes.
Clearly both descriptions are statistically identical.
Here $L=200$, and the quadratic fitness is used with $k = 4.0$, $\mu=1$,
and for a population of $N=10^9$ individuals.
\label{fig7}
}
\end{figure}

As a final remark, we tested the validity of the description of the stochastic
process in the language of Hamming distance classes, as used in our theory.
For that purpose, we performed numerical experiments with Lebowitz-Gillespie
simulations with both a finite population of explicit sequences
\cite{Munoz08}, and
the analogous system in the representation of Hamming distance classes.
As expected from a simple argument based on permutation invariance of
the fitness function that shows the stochastic class dynamics
is an exact projection of the stochastic sequence dynamics, 
both descriptions yield exactly the same statistics,
as shown in Fig.\ \ref{fig7}.  

\section{The Eigen model}

We now turn to the Eigen model.
In contrast to the parallel model,
mutation and horizontal gene transfer 
are assumed to occur only during replication in the Eigen model.
That is, multiple mutations occur along each
sequence as a consequence of errors in the replication process,
and during this process horizontal gene transfer with probability
$\nu/L$ per letter can also occur.
The transfer matrix
for mutations from class $\xi^{'}$ into class $\xi$ 
is denoted by $Q_{\xi,\xi^{'}}$ \cite{Park06},
\begin{eqnarray}
Q_{\xi,\xi^{'}}&=& \sum_{\xi_{1} = 0}^{\min\{\xi+\xi^{'},2L - (\xi+\xi^{'})\}}
q^{L - (2 \xi_{1} + |\xi^{'} - \xi|)}\nonumber\\
&\times &(1-q)^{2\xi_{1} + |\xi - \xi^{'}|}
\binom{L - \xi^{'}}{\xi_{1} + \frac{|\xi^{'} - \xi|-\xi^{'}+\xi}{2}}
\nonumber\\
&\times &\binom{\xi^{'}}{\xi_{1} + \frac{|\xi^{'}-\xi| + \xi^{'}-\xi}{2}}
\label{eq21}
\end{eqnarray}
Here, $q \simeq 1$ characterizes the fidelity in the replication process,
when $1 - q$ is the probability (per site) that an incorrect letter
is placed by the polymerase enzyme. Note that 'back mutations', often
ignored in the literature, are included in the Eigen model.
There is also random degradation of individuals with rate $L d$.
We again seek to calculate shifts in the average population
distribution as well as fluctuations about the average
for a finite population of individuals following the
dynamics of the Eigen model master equation. Here, terms
proportional to $(1-\nu/L)$ represents the evolutionary
processes of replication and multiple mutations in the absence of
horizontal gene transfer. On the other hand, the terms
proportional to $\nu/L$ represent the coupled sequential processes
of replication, horizontal gene transfer and multiple mutations. We also
consider the possibility of degradation through terms proportional to
the degradation rate $d(\xi)$.
\begin{eqnarray}
\frac{\partial}{\partial t}P(\{n_{\xi}\})&=&
\bigg(1-\frac{\nu}{L}\bigg)\bigg\{
\sum_{\xi=0}^{L}r(\xi)Q_{\xi,\xi}\bigg[(n_{\xi}-1)\nonumber\\
&\times &\sum_{\xi^{''}\ne\xi}
\frac{n_{\xi^{''}}+1}{N}
P(n_{\xi}-1,n_{\xi^{''}}+1)\nonumber\\
&-& n_{\xi}\sum_{\xi^{''}\ne\xi}\frac{n_{\xi^{''}}}{N}
P(n_{\xi},n_{\xi^{''}}) \bigg]
+\sum_{\xi=0}^{L}r(\xi)\nonumber\\
&\times &\sum_{\xi^{'}\ne\xi}Q_{\xi^{'},\xi}
\bigg[n_{\xi}\frac{n_{\xi}+1}{N}P(n_{\xi}+1,n_{\xi^{'}}-1)\nonumber\\
&-&(n_{\xi}-1)\frac{n_{\xi}}{N}P(n_{\xi},n_{\xi^{'}}) \bigg]
+ \sum_{\xi=0}^{L}r(\xi)\nonumber\\
&\times &\sum_{\xi^{'}\ne\xi}Q_{\xi^{'},\xi}
\bigg[n_{\xi}\sum_{(\xi^{''}\ne\xi,\xi^{''}\ne\xi^{'})}
\frac{n_{\xi^{''}}+1}{N}\nonumber\\
&\times & P(n_{\xi^{'}}-1,n_{\xi^{''}}+1)
-n_{\xi}\sum_{(\xi^{''}\ne\xi,\xi^{''}\ne\xi^{'})}\frac{n_{\xi^{''}}}{N}
\nonumber\\
&\times &P(n_{\xi^{'}},n_{\xi^{''}}) \bigg]\bigg\}
+\sum_{\xi=0}^{L}d(\xi)\bigg[(n_{\xi}+1)\nonumber\\
&\times &\sum_{\xi^{'}\ne\xi}
\frac{n_{\xi^{'}}-1}{N}P(n_{\xi}+1,n_{\xi^{'}}-1)
-n_{\xi}\nonumber\\
&\times &\sum_{\xi^{'}\ne\xi}\frac{n_{\xi^{'}}}{N}
P(n_{\xi},n_{\xi^{'}}) \bigg]
+ \sum_{\xi,\xi^{'}=0}^{L}Q_{\xi^{'},\xi+1}\frac{\nu}{L}\rho_{+}
\nonumber\\
&\times &(L-\xi)r(\xi)n_{\xi}\sum_{(\xi^{''}\ne\xi,\xi^{''}\ne\xi^{'})}
\bigg[\frac{n_{\xi^{''}}+1}{N}\nonumber\\
&\times&P(n_{\xi^{'}}-1,n_{\xi^{''}}+1)-
\frac{n_{\xi^{''}}}{N}P(n_{\xi^{'}},n_{\xi^{''}}) \bigg]\nonumber\\
&+& \sum_{\xi,\xi^{'}=0}^{L}Q_{\xi^{'},\xi-1}
\frac{\nu}{L}\rho_{-}\xi r(\xi)
n_{\xi}\nonumber\\
&\times&\sum_{(\xi^{''}\ne\xi,\xi^{''}\ne\xi^{'})}
\bigg[\frac{n_{\xi^{''}}+1}{N}P(n_{\xi^{'}}-1,n_{\xi^{''}}+1)
\nonumber\\
&-& \frac{n_{\xi^{''}}}{N}P(n_{\xi^{'}},n_{\xi^{''}}) \bigg]
\nonumber\\
\label{eq22}
\end{eqnarray}

\subsection{Mapping to a field theory}
By the same method as in the parallel model, we map the master
equation into a second quantized formulation, with Hamiltonian
\begin{eqnarray}
-\hat{H} 
&=& \bigg(1-\frac{\nu}{L} \bigg)
(L/N) \sum_{\xi,\xi^{'},\xi^{''}=0}^{L}
Q_{\xi^{'},\xi} f(\xi)\hat{a}_{\xi}^{\dagger}(\hat{a}_{\xi^{'}}^{\dagger}
-\hat{a}_{\xi^{''}}^{\dagger})\nonumber\\
&\times&\hat{a}_{\xi}\hat{a}_{\xi^{''}}
+(L/N) \sum_{\xi,\xi^{'}=0}^{L}  d(\xi^{'})\hat{a}_{\xi}^{\dagger}
(\hat{a}_{\xi}^{\dagger}-\hat{a}_{\xi^{'}}^{\dagger})\hat{a}_{\xi}
\hat{a}_{\xi^{'}}
\nonumber\\ &+&
(L/N) \sum_{\xi,\xi^{'},\xi^{''}=0}^{L}Q_{\xi^{'},\xi+1}
(\nu/L) 
\rho_{+}(L-\xi) f(\xi)\hat{a}_{\xi}^{\dagger}(\hat{a}_{\xi^{'}}^{\dagger}
\nonumber\\
&-&
\hat{a}_{\xi^{''}}^{\dagger})\hat{a}_{\xi}\hat{a}_{\xi^{''}}
+ (L/N) \sum_{\xi,\xi^{'},\xi^{''}=0}^{L}Q_{\xi^{'},\xi-1}
(\nu/L) \rho_{-}\xi\nonumber\\
&\times &  f(\xi)\hat{a}_{\xi}^{\dagger}(
\hat{a}_{\xi^{'}}^{\dagger}-\hat{a}_{\xi^{''}}^{\dagger})
\hat{a}_{\xi}\hat{a}_{\xi^{''}}
\label{eq23}
\end{eqnarray}
With a similar method as in the parallel model, we introduce coherent states
in a Trotter factorization of the evolution operator, as defined in 
Eq.\ (\ref{eq8}). 
From this procedure, we derive the field theory for the Eigen model as well. 
In this case, the action given by
\begin{widetext}
\begin{eqnarray}
&&S[\{z\},\{\bar{z}\}] = 
\sum_{\xi=0}^{L}\int_{0}^{T}dt^{'}\bigg\{\bar{z}_{\xi}\frac{\partial z_{\xi}}
{\partial t'}+ \bigg(\bar{z}_{\xi}(t')
z_{\xi}(t')-n_{\xi}^{0}\ln[1 + \bar{z}_{\xi}(t')]\bigg)\delta(t^{'})
-\frac{L}{N}\bigg(1-\frac{\nu}{L}\bigg) \sum_{\xi^{'},\xi^{''}=0}^{L}
Q_{\xi^{'},\xi} f(\xi)[1+\bar{z}_{\xi}]\nonumber\\
&&\times[\bar{z}_{\xi^{'}}
-\bar{z}_{\xi^{''}}]z_{\xi}z_{\xi^{''}}
-\frac{L}{N} 
\sum_{\xi^{'},\xi^{''}=0}^{L}\bigg[\delta_{\xi,\xi^{'}} d(\xi^{''})+
\frac{\nu}{L}[Q_{\xi^{'},\xi+1}\rho_{+}(L-\xi)
+ Q_{\xi^{'},\xi-1}\rho_{-}\xi] f(\xi)\bigg][1+\bar{z}_{\xi}]
[\bar{z}_{\xi^{'}}-\bar{z}_{\xi^{''}}]z_{\xi}z_{\xi^{''}}\bigg\}
\label{eq24}
\end{eqnarray}
\end{widetext}

In the limit of a large population, 
we look for a saddle-point in the action Eq.\ (\ref{eq24}). From the
condition $\left.\frac{\delta S}{\delta z_{\xi}(t)}\right|_{c}=0$, we
obtain $\bar{z}_{\xi}^{c}(t)=0$.
From the second equation
$\left.\frac{\delta S}{\delta \bar{z}_{\xi}(t)}\right|_{c}=0$, we find that
$P_{\xi}(t) = z_{\xi}^{c}(t)/N$ satisfies the differential equation
\begin{eqnarray}
\frac{d}{dt}P_{\xi}(t) &=& \left(1-\frac{\nu}{L}\right)
\bigg[\sum_{\xi^{'}=0}^{L}Q_{\xi,\xi^{'}}r(\xi^{'})P_{\xi^{'}}(t)-P_{\xi}(t)
\nonumber\\
&\times &\sum_{\xi^{'}=0}^{L}r(\xi^{'})P_{\xi^{'}}(t)\bigg]
- P_{\xi}(t)\bigg[d(\xi) - \sum_{\xi^{'}=0}^{L}P_{\xi^{'}}(t)\nonumber\\
&\times & d(\xi^{'})\bigg]
+\frac{\nu}{L}\bigg[\sum_{\xi^{'}=0}^{L}\bigg\{Q_{\xi,\xi^{'}+1}\rho_{+}
(L-\xi^{'})\nonumber\\
&+ & Q_{\xi,\xi^{'}-1}\rho_{-}\xi^{'}\bigg\}r(\xi^{'})P_{\xi^{'}}(t)
-P_{\xi}(t)\nonumber\\
&\times&\sum_{\xi^{'}=0}^{L}
\{\rho_{+}(L-\xi^{'})+\rho_{-}\xi^{'}\}r(\xi^{'})
P_{\xi^{'}}(t)\bigg]
\label{eq25}
\end{eqnarray}
and the initial condition corresponds
to $P_{\xi}(0)=n_{\xi}^{0}/N$, as derived
in Appendix 3. 
This is exactly the differential
equation  
for $P_{\xi}(t)$ from infinite population quasispecies theory
\cite{Park07,Munoz08}.

\begin{figure}[tbp]
\centering
\epsfig{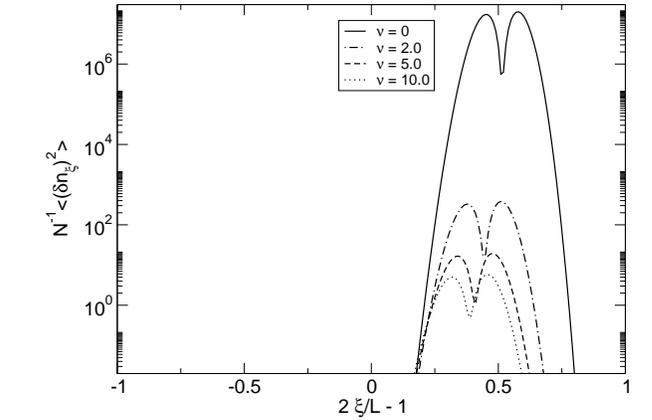}
\caption{Fluctuations in the probability distribution,
as predicted from our theory
Eqs.\ (\ref{eq25}--\ref{eq26}), for the Eigen model and quadratic fitness,
at different horizontal gene transfer rates, $\nu$. Here $L=200$, $k=4.0$,
and $\mu=1$.
Fluctuations decrease by orders of magnitude with increasing
horizontal gene transfer rate.
\label{fig5}
}
\end{figure}

By expanding the action Eq.\ (\ref{eq24}) up to second order to
calculate the matrix of correlations, as shown in
Appendix 4, we obtain 
in the continuous time limit
the Lyapunov Eq.\ (\ref{eq12}), with matrices $A$
defined by 
\begin{eqnarray}
&& L^{-1} [A]_{\xi,\xi^{'}} = \bigg(1-\frac{\nu}{L}\bigg)
\bigg[\sum_{\xi^{''}=0}^{L}
Q_{\xi,\xi^{''}}f(\xi^{''})P_{\xi^{''}}
\nonumber\\
&&+ Q_{\xi,\xi^{'}}f(\xi^{'})-\delta_{\xi,\xi^{'}}\sum_{\xi^{''}=0}^{L}f(\xi^{''})P_{\xi^{''}}
-f(\xi^{'})P_{\xi}\bigg]\nonumber\\
&&+[d(\xi^{'})-d(\xi)]P_{\xi}
+ \delta_{\xi,\xi^{'}}\bigg[
\sum_{\xi_{1}=0}^{L}d(\xi_{1})P_{\xi_{1}} - d(\xi)\bigg]
\nonumber\\
&&+\frac{\nu}{L}\bigg[\sum_{\xi^{''}=0}^{L}\bigg(Q_{\xi,\xi^{''}-1}
\rho_{-}\xi^{''}+ Q_{\xi,\xi^{''}+1}\rho_{+}(L-\xi^{''})\bigg)
\nonumber\\
&&\times f(\xi^{''})
P_{\xi^{''}} + \bigg(Q_{\xi,\xi^{'}-1}\rho_{-}
\xi^{'}
+Q_{\xi,\xi^{'}+1}\rho_{+}(L-\xi^{'})  \bigg)\nonumber\\
&&\times f(\xi^{'})
-\delta_{\xi,\xi^{'}}\sum_{\xi^{''}=0}^{L}\bigg(\rho_{+}(L-\xi^{''})
+\rho_{-}\xi^{''}  \bigg)\nonumber\\
&&\times f(\xi^{''})P_{\xi^{''}}
-\bigg(\rho_{+}(L-\xi^{'}) + \rho_{-}\xi^{'}  \bigg)f(\xi^{'})P_{\xi}
\bigg]
\label{eq26}
\end{eqnarray}
and matrices $B$ given by
\begin{eqnarray}
&&L^{-1} [B]_{\xi,\xi^{'}} =N\bigg\{\bigg(1-\frac{\nu}{L} \bigg)
\bigg[Q_{\xi^{'},\xi}
f(\xi) P_{\xi}\nonumber\\
&&+ Q_{\xi,\xi^{'}} f(\xi^{'}) P_{\xi^{'}}
-(f(\xi)+ f(\xi^{'}))P_{\xi}P_{\xi^{'}} \bigg]
\nonumber\\
&&+2\bigg(\sum_{\xi_{1}=0}^{L}d(\xi_{1})P_{\xi_{1}}\bigg)
P_{\xi}
\delta_{\xi,\xi^{'}}
+\frac{\nu}{L}\bigg[\bigg(Q_{\xi^{'},\xi+1}\rho_{+}(L-\xi)
\nonumber\\
&&+ Q_{\xi^{'},\xi-1}\rho_{-}\xi \bigg)
f(\xi) P_{\xi}
+\bigg(Q_{\xi,\xi^{'}+1}\rho_{+}(L-\xi^{'})
\nonumber\\
&&+ Q_{\xi,\xi^{'}-1}\rho_{-}
\xi^{'} \bigg)f(\xi^{'}) P_{\xi^{'}}
-\bigg[\bigg(\rho_{+}(L-\xi) + \rho_{-}\xi \bigg)f(\xi) 
\nonumber\\
&&+\bigg(\rho_{+}(L-\xi^{'})
+\rho_{-}\xi^{'}\bigg)f(\xi^{'})\bigg]P_{\xi}P_{\xi^{'}}
\nonumber\\
&&-(d(\xi)
+ d(\xi^{'}))P_{\xi} P_{\xi^{'}}
\bigg\}
\label{eq27}
\end{eqnarray}

\begin{figure}[tbp]
\centering
\epsfig{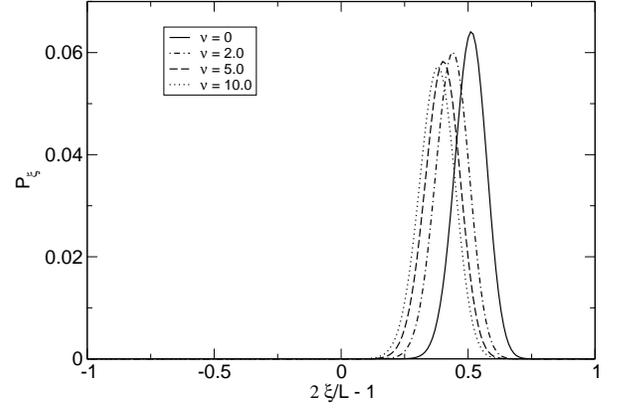}
\caption{Probability distributions,
as predicted from our theory, for the Eigen model and quadratic fitness,
at different recombination rates. Here $L=200$, $k = 4.0$, and $\mu=1$.
\label{fig6}
}
\end{figure}

\subsection{Continuous and discontinuous fitness functions}
For the sharp peak $f(\xi) = (A-A_{0})\delta_{\xi,L} + A_{0}$, 
for the Eigen model in the absence of horizontal gene transfer
($\nu = 0$), we obtain that the wild type probability
is
\begin{eqnarray}
\sum_{\xi^{'}=0}^{L}q^{\xi^{'}}(1-q)^{L-\xi^{'}}f(\xi^{'})P_{\xi^{'}}
&-&P_{L}[A P_{L} \nonumber\\
&+& A_{0}\sum_{\xi^{'}\ne L}P_{\xi^{'}}]=0
\label{eq28}
\end{eqnarray}
Since $q \simeq 1$, (the fidelity in the replication process is very high),
then $1 - q \ll 1$ and Eq.\ (\ref{eq28}) becomes.
\begin{eqnarray}
q^{L} A P_{L} - P_{L}[(A - A_{0})P_{L} + A_{0}] = 0
\label{eqA29}
\end{eqnarray}
By defining $q^{L}=e^{-\mu}$, we obtain for the probability of
the wild-type
\begin{eqnarray}
P_{\xi=L} = \left\{\begin{array}{cc} 0, & A < e^{\mu}A_{0}\\
(e^{-\mu}A-A_{0})/(A-A_{0}), & A > e^{\mu}A_{0}\end{array}\right.
\label{eq30}
\end{eqnarray}
For the correlation matrix, we define $D_{\xi,\xi^{'}} =
\frac{1}{N}C_{\xi,\xi^{'}}$, and find that the stationary solution
for $D_{L,L}$ in the absence of degradation $d(\xi) = 0$ is given
by
\begin{eqnarray}
0 = \frac{1}{N}B_{L,L} + \sum_{\xi_{1}=0}^{L}[A_{L,\xi_{1}}D_{\xi_{1},L}
+ A_{L,\xi_{1}}D_{\xi_{1},L}]
\label{eq31}
\end{eqnarray}
From this equation, we find
$\sum_{\xi_{1}}A_{L,\xi_{1}}D_{\xi_{1},L} = -\frac{1}{2N}B_{L,L}$. Hence,
expanding the left hand side explicitly, we find
\begin{eqnarray}
\sum_{\xi_{1}=0}^{L}\bigg[&&\sum_{\xi^{''}=0}^{L}
Q_{L,\xi^{''}}f(\xi^{''})P_{\xi^{''}} + Q_{L,\xi_{1}}f(\xi_{1})
\nonumber\\
&-&(\sum_{\xi_{1}^{'}}f(\xi_{1}^{'})P_{\xi_{1}^{'}})\delta_{L,\xi_{1}}
-f(\xi_{1})P_{L} \bigg]D_{\xi_{1},L}\nonumber\\
&=&-[Q_{L,L}f(L)P_{L} - f(L)P_{L}^{2} ]
\label{eq32}
\end{eqnarray}
Expanding this equation when $L$ is large and $q\simeq 1$, we find
\begin{eqnarray}
[q^{L} A &-& (A - A_{0})P_{L} - A_{0} - (A - A_{0})P_{L} ]D_{L,L}
\nonumber\\
&=& A P_{L}(P_{L} - q^{L}) + q^{L}A P_{L}^{2} - A_{0}P_{L}^{2}
\label{eq33}
\end{eqnarray}
Substituting the result $P_{L} = \frac{q^{L}A - A_{0}}{A - A_{0}}$
from Eq.\ (\ref{eq30}), we find
\begin{eqnarray}
D_{L,L} = \frac{1}{(A - A_{0})^{2}}[A A_{0} - A_{0}^{2} - (q^{L}A)^{2}+
q^{L}A A_{0} ]
\label{eq34}
\end{eqnarray}
The fluctuation in the number of individuals with the wild-type
sequence is obtained from Eq.\ (\ref{eq15}),
\begin{eqnarray}
\frac{\langle(\delta n_{\xi=L})^{2}\rangle}{N^{2}} 
= \left\{\begin{array}{cc}0, & A < e^{\mu}A_{0}\\
\frac{e^{-\mu}(1-e^{-\mu})A^{2}}{N(A-A_{0})^{2}}, & A > e^{\mu}A_{0}\end{array}\right.
\label{eq36}
\end{eqnarray}

For smooth fitness functions, there are large fluctuations in the
population numbers in the absence of horizontal gene transfer.
In Fig.\ \ref{fig5} we present the fluctuations in the
number of individuals with a given sequence for the quadratic fitness, as
predicted from our theory Eqs.\ (\ref{eq25}--\ref{eq27}). 
A moderate horizontal gene
transfer rate reduces by orders of magnitude the fluctuations.
In Fig.\ \ref{fig6} inset, we present the equilibrium probability
distributions, for different rates of horizontal gene transfer,
as obtained from our theory for the
quadratic fitness $f(\xi)=(k/2) (2\xi/L-1)^2/2 + 1$. 
For this fitness function with negative epistasis, 
horizontal gene transfer reduces the mean fitness in the
infinite population limit \cite{Munoz08}.

\section{Conclusion}
For both the parallel and Eigen models, we have found that horizontal
gene transfer reduces by orders
of magnitude the fluctuations in the number of individuals
with a given sequence composition
for smooth fitness functions, such as quadratic.
Horizontal gene transfer
also reduces the variability within and
between independent experiments for
smooth fitness functions.
Finally, horizontal
gene transfer substantially reduces the ``Muller's ratchet''
phenomenon, whereby fitness is reduced in finite populations
relative to the infinite population limit.
For the sharp peak fitness, horizontal gene transfer
does not modify the steady-state distribution of fluctuations.

The reduction in finite populations by horizontal gene transfer of both the
magnitude  of the Muller's ratchet  phenomenon
\cite{Chao92,Chao97,Rice01}
and the fluctuations in population numbers
should be observable in experiments.  The fluctuation in population
numbers can be measured either at different time points
in long experiments or as fluctuations
between different experimental replicates.  The latter is likely
to be more feasible in the laboratory.

\section{Acknowledgments}
Supported by the FunBio program of DARPA.
JMP is also supported by a Korea Research Foundation grant funded
by the Korean Government (KRF-2008-314-C00123).

\section{Appendix 1}

We present the derivation of the saddle point equations for the Kimura model.
We look for a saddle point of the action Eq.\ (\ref{eq10}) 
in the coherent fields
$z_{\xi}(t)$ and $\bar{z}_{\xi}(t)$. The first condition is
\begin{eqnarray}
\frac{\delta S}{\delta z_{\xi}(t)} &=& -\frac{\partial\bar{z}_{\xi}}
{\partial t}+\delta(t-T)\bar{z}_{\xi}(T) 
- \mu[(L-\xi)\bar{z}_{\xi+1}(t) \nonumber\\
&+& \xi\bar{z}_{\xi-1}(t)-L\bar{z}_{\xi}(t)]
-\nu[(L-\xi)\rho_{+}\bar{z}_{\xi+1}(t)\nonumber\\
&+&\xi\rho_{-}\bar{z}_{\xi-1}
-\{(L-\xi)\rho_{+} + \xi\rho_{-}\}\bar{z}_{\xi}(t)]\nonumber\\
&-&\frac{1}{N}\sum_{\xi_{1}=0}^{L}\sum_{\xi_{2}=0}^{L} L f(\xi_{1})
(1 + \bar{z}_{\xi_{1}}(t))[\bar{z}_{\xi_{1}}(t)
\nonumber\\
&-&\bar{z}_{\xi_{2}}(t)](\delta_{\xi_{1},\xi}z_{\xi_{2}}(t)
+ z_{\xi_{1}}(t)\delta_{\xi_{2},\xi})=0
\label{eqA1}
\end{eqnarray}
where $T$ is the final integration time in Eq.\ (\ref{eq10}), which
we typically set as $T = \infty$.
The solution which satisfies this saddle-point condition is 
$\bar{z}_{\xi}^{c}(t)=0$, for $0 < t < T$.

The saddle point condition in the fields $\bar{z}_{\xi}(t)$ is
\begin{eqnarray}
\frac{\delta S}{\delta \bar{z}_{\xi}(t)} &=& [z_{\xi}(0) - \frac{n_{\xi}(0)}
{1+\bar{z}_{\xi}(0)}]\delta(t) + \frac{\partial z_{\xi}}{\partial t}
-\mu[ (L-\xi+1)\nonumber\\
&\times & z_{\xi-1}(t)+(\xi+1)z_{\xi+1}(t)-L z_{\xi}(t)]
-\nu[(L-\xi\nonumber\\
&+& 1)\rho_{+}z_{\xi-1}(t)+(\xi+1)\rho_{-}z_{\xi+1}(t)
-\{(L-\xi)\rho_{+}\nonumber\\
&+& \xi\rho_{-}\}z_{\xi}(t)]
-\frac{1}{N}\sum_{\xi_{1}=0}^{L}\sum_{\xi_{2}=0}^{L}L f(\xi_{1})\{
\delta_{\xi_{1},\xi}
[\bar{z}_{\xi_{1}}(t)\nonumber\\ 
&-& \bar{z}_{\xi_{2}}(t)] 
+ (1+\bar{z}_{\xi_{1}})[\delta_{\xi_{1},\xi}-\delta_{\xi_{2},\xi}]\}
z_{\xi_{1}}(t)z_{\xi_{2}}(t)=0
\nonumber\\
\label{eqA2}
\end{eqnarray}
In combination with the solution $\bar{z}_{\xi}^{c}(t)=0$ obtained from 
Eq.\ (\ref{eqA1}), Eq.\ (\ref{eqA2}) provides the differential equation
for the probability distribution $P_{\xi}(t)=z_{\xi}^{c}(t)/N$,
\begin{eqnarray}
\frac{d}{dt}P_{\xi} &=& \mu[(L-\xi+1)P_{\xi-1}+(\xi+1)P_{\xi+1}-L P_{\xi}]
\nonumber\\
&+&\nu[\rho_{+}(L-\xi+1)P_{\xi-1}+\rho_{-}(\xi+1)P_{\xi+1}\nonumber\\
&-&
\{(L-\xi)\rho_{+}+\xi\rho_{-} \}P_{\xi}]
+ [r(\xi) - \sum_{\xi'=0}^{L}r(\xi')P_{\xi'}]P_{\xi}\nonumber\\
\label{eqA3}
\end{eqnarray} 
and the initial condition $P_{\xi}(0)=n_{\xi}^{0}/N$. 
In deriving Eq.\ (\ref{eqA3}) from Eq.\ (\ref{eqA2}), the property 
$\sum_{\xi=0}^{L}P_{\xi}(t)=1$ was used, 
and we introduce the notation $r(\xi)=L f(\xi)$.

\section{Appendix 2}

We next consider the expansion of the action Eq.\ (\ref{eq10})
near the saddle-point
$S_{c}$. For convenience, we define a discrete time label $k=t/\epsilon$,
with $\epsilon\rightarrow 0$. Fluctuations near the saddle-point
solution are given by $\delta z_{\xi}(k) = z_{\xi}(k) - z_{\xi}^{c}(k)$,
and $\delta \bar{z}_{\xi}(k) = \bar{z}_{\xi}(k) - \bar{z}_{\xi}^{c}(k)$.
This gives
\begin{eqnarray}
S - S_{c} &=& \sum_{\xi,\xi^{'}=0}^{L}\bigg[\delta\bar{z}_{\xi}(0)
\delta z_{\xi^{'}}(0)\delta_{\xi,\xi^{'}}+\frac{1}{2}n_{\xi}^{0}
\delta\bar{z}_{\xi}(0)\delta\bar{z}_{\xi^{'}}(0)\nonumber\\
&\times &\delta_{\xi,\xi^{'}}
+\sum_{k=1}^{t/\epsilon}\bigg\{\delta\bar{z}_{\xi}(k)\delta z_{\xi^{'}}(k)
\delta_{\xi,\xi^{'}}-\epsilon\delta\bar{z}_{\xi}(k)\delta\bar{z}_{\xi^{'}}(k)
\nonumber\\
&\times &[\delta_{\xi,\xi^{'}}r(\xi)N P_{\xi}(k-1)
- r(\xi)N P_{\xi}(k-1)\nonumber\\
&\times & P_{\xi^{'}}(k-1)]\bigg\}
+\sum_{k=1}^{t/\epsilon}\delta\bar{z}_{\xi}(k)\delta z_{\xi^{'}}(k-1)
\bigg\{-\delta_{\xi,\xi^{'}}\nonumber\\
&-&\epsilon\mu[(L-\xi+1)\delta_{\xi-1,\xi^{'}}
+(\xi+1)\delta_{\xi+1,\xi^{'}}-L\delta_{\xi,\xi^{'}} ]
\nonumber\\
&-&\epsilon\nu[(L-\xi+1)\rho_{+}\delta_{\xi-1,\xi^{'}}
+(\xi+1)\rho_{-}\delta_{\xi+1,\xi^{'}}\nonumber\\
&-&\{(L-\xi)\rho_{+}+\xi\rho_{-}\}
\delta_{\xi,\xi^{'}} ]- \epsilon[\{r(\xi)-\sum_{\xi_{1}}r(\xi_{1})
\nonumber\\
&\times& P_{\xi_{1}}(k-1)\}
\delta_{\xi,\xi^{'}}+(r(\xi)
-r(\xi^{'}))P_{\xi}(k-1)]\bigg\}\bigg]\nonumber\\
&=&\frac{1}{2} X^{T}\Pi^{-1} X + O(X^{3})
\label{eqA4}
\end{eqnarray}
Here, we have defined the vector
$X^{T} = \left(\{\delta\bar{z}(0),\delta z(0)\},\ldots,
\{\delta\bar{z}(t/\epsilon),\delta z(t/\epsilon)\} \right)$.
The matrix $\Pi^{-1}$ is banded tri-diagonal, with
\begin{eqnarray}
\Pi^{-1}= \left(\begin{array}{cccccc}\Pi_{00}^{-1}&-\Pi_{01}^{-1}&0&0&\ldots&0\\
-\Pi_{10}^{-1}&\Pi_{11}^{-1}&-\Pi_{12}^{-1}&0&\ldots&0\\
0&-\Pi_{21}^{-1}&\Pi_{22}^{-1}&-\Pi_{23}^{-1}&\ldots&0\\
\vdots & & \ddots & & &\vdots \\
\ldots & & & & \ldots & \Pi_{t/\epsilon,t/\epsilon}^{-1} \end{array}\right)
\nonumber\\
\label{eqA5}
\end{eqnarray}
Here,
\begin{eqnarray}
\Pi_{00}^{-1}&=&\left(\begin{array}{cc}N^{0} & I\\I & 0\end{array} \right),
\;\;\;[N^{0}]_{\xi,\xi^{'}}=n_{\xi}^{0}\delta_{\xi,\xi^{'}}\nonumber\\
\Pi_{k,k}^{-1} &=& \left(\begin{array}{cc}-\epsilon B(k-1)&I\\I&0\end{array}
\right), \;\;\;k\ne 0
\nonumber\\
\Pi_{k,k-1}^{-1} &=& \left(\begin{array}{cc}0 & I + \epsilon A(k-1)\\
0 & 0\end{array} \right)\nonumber\\
\Pi_{k-1,k}^{-1} &=& \left(\begin{array}{cc}0&0\\I + \epsilon A^{T}(k-1)
& 0\end{array} \right)
\label{eqA6}
\end{eqnarray}
The matrices $A$ and $B$ are defined by
\begin{eqnarray}
&&[A]_{\xi,\xi^{'}}= \delta_{\xi-1,\xi^{'}}(L-\xi+1)[\mu+\nu\rho_{+}]
+\delta_{\xi,\xi^{'}}[L f(\xi)
\nonumber\\
&&
-
\sum_{\xi_{1}}L f(\xi_{1})P_{\xi_{1}}
-\nu\{(L-\xi)\rho_{+}+\xi\rho_{-}\}
-L\mu]
\nonumber\\
&&
+L [f(\xi)-f(\xi^{'})]P_{\xi}+
\delta_{\xi+1,\xi^{'}}(\xi+1)[\mu+\nu\rho_{-}]
\label{eqA7}
\end{eqnarray}
\begin{eqnarray}
[B]_{\xi,\xi^{'}}=\delta_{\xi,\xi^{'}} 2 L f(\xi) N P_{\xi}
-L [f(\xi)+f(\xi^{'})]N P_{\xi} P_{\xi^{'}}
\label{eqA9}
\end{eqnarray}
Here, $A$  a symmetric matrix $[A^{T}(k)]_{\xi,\xi^{'}}= [A(k)]_{\xi^{'},\xi}$.
By standard matrix inversion, we obtain
\begin{eqnarray}
\Pi(t/\epsilon) &=& \bigg[\Pi^{-1}(t/\epsilon)\bigg]^{-1}\nonumber\\
&=&
\left[\begin{array}{cc}\bigg[\Pi^{-1}(t/\epsilon-1)\bigg]&
\left(\begin{array}{c}0\\0\\ \vdots\\-\Pi^{-1}_{t/\epsilon-1,t/\epsilon}
\end{array}\right)\\\bigg(0\; 0 \ldots -\Pi^{-1}_{t/\epsilon,t/\epsilon-1}
\bigg)&\Pi^{-1}_{t/\epsilon,t/\epsilon} \end{array} \right]^{-1}
\nonumber\\
\label{eqA10}
\end{eqnarray}

Calculating the inverse in Eq.\ (\ref{eqA10}), we obtain
\begin{eqnarray}
\bigg[\Pi(t/\epsilon)\bigg]_{t/\epsilon,t/\epsilon} &\equiv & b_{t/\epsilon,t/\epsilon}
\nonumber\\
&=&\bigg[\Pi^{-1}_{t/\epsilon,t/\epsilon}-\bigg(0\;0\ldots -\Pi^{-1}_{t/\epsilon,t/\epsilon-1}\bigg)
\nonumber\\
&\times &\bigg[\Pi(t/\epsilon-1)\bigg]
\left(\begin{array}{c}0\\0\\ \vdots\\-\Pi^{-1}_{t/\epsilon-1,t/\epsilon} \end{array} \right) \bigg]^{-1}
\nonumber\\
&=&\bigg[\Pi^{-1}_{t/\epsilon,t/\epsilon}-\Pi^{-1}_{t/\epsilon,t/\epsilon-1}
b_{t/\epsilon-1,t/\epsilon-1}\nonumber\\
&\times &\Pi^{-1}_{t/\epsilon-1,t/\epsilon}
\bigg]^{-1}
\label{eqA11}
\end{eqnarray}

From this recursive equation, we find
\begin{eqnarray}
b_{00}&=&\bigg[\Pi^{-1}_{00}\bigg]^{-1}=\left(\begin{array}{cc}
0 & I\\I& -N^{0}\end{array} \right)\nonumber\\b_{11} &=& 
\bigg[\Pi_{11}^{-1}-\Pi_{10}^{-1}\,b_{00}\,\Pi_{01}^{-1} \bigg]^{-1}
\nonumber\\&=&\left[\begin{array}{cc}0& I\\I &\;\;\; 
\{I+\epsilon A(0)\}[-N^{0}]\{I+\epsilon A^{T}(0)\}+\epsilon B(0) 
\end{array} \right]
\nonumber\\
\label{eqA12}
\end{eqnarray}

From Eq.\ (\ref{eqA12}), proceeding by induction, we prove that
the matrices $b_{k}$ possess the structure
\begin{eqnarray}
b_{k,k} = \left(\begin{array}{cc}0 & I\\I & C(k) \end{array} \right),
\label{eqA1_16}
\end{eqnarray}
and after the recursion relation
\begin{eqnarray}b_{k,k} = \bigg[\Pi^{-1}_{k} - \Pi^{-1}_{k,k-1} b_{k-1,k-1}\Pi^{-1}_{k-1,k} \bigg]^{-1}\label{eqA1_17}
\label{eqA13}
\end{eqnarray}
we obtain
\begin{eqnarray}
C(k) &=& [I + \epsilon A(k-1)]C(k-1)[I + \epsilon A^{T}(k-1)] \nonumber\\
&+ &\epsilon B(k-1)\nonumber\\
C(0) &=& -N^{0}
\label{eqA14}
\end{eqnarray}

In the continuous time limit $\epsilon\rightarrow 0$, 
Eq.\ (\ref{eqA14}) becomes a
Lyapunov equation
\begin{eqnarray}
\frac{d}{dt}C &=& B + A C + C A^{T}\nonumber\\C(0) &=& - N^{0}
\label{eqA15}
\end{eqnarray}
with $[N^{0}]_{\xi,\xi^{'}} = \delta_{\xi,\xi^{'}}n_{\xi}^{0}$.

\section{Appendix 3}

Now, we derive the saddle point equations for the Eigen model.
We look for a saddle point of the action Eq.\ (\ref{eq24}) 
in the coherent fields
$z_{\xi}(t)$ and $\bar{z}_{\xi}(t)$. The first condition is
\begin{eqnarray}
\frac{\delta S}{\delta z_{\xi}(t)}&=&-\frac{\partial\bar{z}_{\xi}}{\partial t}
+\delta(t-T)\bar{z}_{\xi}(T)
-\frac{L}{N}\bigg(1-\frac{\nu}{L}\bigg)\nonumber\\
&\times &\sum_{\xi_{1},\xi_{2},\xi_{3} = 0}^{L}
\bigg\{Q_{\xi_{2},\xi_{1}}
f(\xi_{1})[1+\bar{z}_{\xi_{1}}(t)][\bar{z}_{\xi_{2}}(t)
\nonumber\\
&-&\bar{z}_{\xi_{3}}(t)]
(\delta_{\xi_{1},\xi}z_{\xi_{3}}(t)
+ \delta_{\xi_{3},\xi}z_{\xi_{1}}(t))\bigg\}\nonumber\\
&-& \frac{L}{N}\sum_{\xi_{1},\xi_{2},\xi_{3}=0}^{L}\bigg\{
\bigg[\delta_{\xi_{1},\xi_{2}}d(\xi_{3})
\nonumber\\
&+&\frac{\nu}{L}[Q_{\xi_{2},\xi_{1}+1}\rho_{+}(L-\xi_{1})
+ Q_{\xi_{2},\xi_{1}-1}\rho_{-}
\xi_{1}]\nonumber\\
&\times &f(\xi_{1})\bigg][1 + \bar{z}_{\xi_{1}}(t)][\bar{z}_{\xi_{2}}(t)
-\bar{z}_{\xi_{3}}(t)](z_{\xi_{3}}(t)\delta_{\xi_{1},\xi}
\nonumber\\
&+&z_{\xi_{1}}(t)\delta_{\xi_{3},\xi})\bigg\}=0
\label{eqA16}
\end{eqnarray}
where $T$ is the total integration time in Eq.\ (\ref{eq24}),
which we typically set as $T=\infty$.
This saddle-point condition is satisfied by the solution 
$\bar{z}_{\xi}^{c}(t)=0$, for $0 < t < T$.

The saddle-point condition in the fields $\bar{z}_{\xi}(t)$ is
\begin{eqnarray}
\frac{\delta S}{\delta \bar{z}_{\xi}(t)}&=&\frac{\partial z_{\xi}}{\partial t}
+\bigg(z_{\xi}(0)-\frac{n^{0}_{\xi}}{1+\bar{z}_{\xi}(0)}
\bigg)\delta(t)-\frac{L}{N}\nonumber\\
&\times &\bigg(1-\frac{\nu}{L}\bigg)
\sum_{\xi_{1},\xi_{2},\xi_{3}=0}^{L}
\bigg\{Q_{\xi_{2},\xi_{1}}f(\xi_{1})
\bigg(\delta_{\xi_{1},\xi}[\bar{z}_{\xi_{2}}(t)
\nonumber\\
&-&\bar{z}_{\xi_{3}}(t)]+[1+\bar{z}_{\xi_{1}}(t)][\delta_{\xi_{2},\xi}
-\delta_{\xi_{3},\xi}]\bigg)z_{\xi_{1}}(t)z_{\xi_{3}}(t)\bigg\}
\nonumber\\
&-&\frac{L}{N}
\sum_{\xi_{1},\xi_{2},\xi_{3}}\bigg\{\bigg[
\delta_{\xi_{1},\xi_{2}}
d(\xi_{3})+\frac{\nu}{L}[Q_{\xi_{2},\xi_{1}+1}\rho_{+}(L-\xi_{1})
\nonumber\\
&+&Q_{\xi_{2},\xi_{1}-1}\rho_{-}\xi_{1}]f(\xi_{1})\bigg]
\bigg[\delta_{\xi_{1},\xi}[\bar{z}_{\xi_{2}}(t)-\bar{z}_{\xi_{3}}(t)]
\nonumber\\
&+&[1+\bar{z}_{\xi_{1}}(t)]
(\delta_{\xi_{2},\xi}-\delta_{\xi_{3},\xi})\bigg]z_{\xi_{1}}(t)
z_{\xi_{3}}(t)\bigg\}=0
\label{eqA17}
\end{eqnarray}

In combination with the solution $\bar{z}_{\xi}^{c}(t)=0$ obtained from
Eq. (\ref{eqA16}), after Eq. (\ref{eqA17}) we obtain the differential
equation for the probability distribution $P_{\xi}(t)=z_{\xi}^{c}(t)/N$,
\begin{eqnarray}
\frac{d}{dt}P_{\xi}(t) &=& \left(1-\frac{\nu}{L}\right)
\bigg[\sum_{\xi^{'}=0}^{L}Q_{\xi,\xi^{'}}r(\xi^{'})P_{\xi^{'}}(t)-P_{\xi}(t)
\nonumber\\
&\times &\sum_{\xi^{'}=0}^{L}r(\xi^{'})P_{\xi^{'}}(t)\bigg]
- P_{\xi}(t)\bigg[d(\xi) - \sum_{\xi^{'}=0}^{L}P_{\xi^{'}}(t)\nonumber\\
&\times &d(\xi^{'})\bigg]
+\frac{\nu}{L}\bigg[\sum_{\xi^{'}=0}^{L}\bigg\{Q_{\xi,\xi^{'}+1}\rho_{+}
(L-\xi^{'})\nonumber\\
&+&Q_{\xi,\xi^{'}-1}\rho_{-}\xi^{'}\bigg\}r(\xi^{'})P_{\xi^{'}}(t)
-P_{\xi}(t)\nonumber\\
&\times&\sum_{\xi^{'}=0}^{L}
\{\rho_{+}(L-\xi^{'})
+\rho_{-}\xi^{'}\}r(\xi^{'})
P_{\xi^{'}}(t)\bigg]
\label{eqA18}
\end{eqnarray}
and the initial condition $P_{\xi}(0)=n_{\xi}^{0}/N$. In deriving
Eq. (\ref{eqA18}) from Eq. (\ref{eqA17}), we used the properties:
$\sum_{\xi=0}^{L}P_{\xi}=1$, and $\sum_{\xi=0}^{L}Q_{\xi,\xi'}=1$.

\section{Appendix 4}

Now, let us consider the expansion of the action Eq.\ ({\ref{eq24})
for the Eigen model near the saddle point, with
fluctuations near the saddle-point
solution given by $\delta z_{\xi}(k) = z_{\xi}(k) - z_{\xi}^{c}(k)$,
and $\delta \bar{z}_{\xi}(k) = \bar{z}_{\xi}(k) - \bar{z}_{\xi}^{c}(k)$. 
\begin{eqnarray}
S - S_{c} &=& \sum_{\xi=0}^{L}\bigg[
\delta\bar{z}_{\xi}(0)\delta z_{\xi}(0) +\frac{1}{2}n_{\xi}^{0}\delta
\bar{z}_{\xi}(0)\delta\bar{z}_{\xi}(0)\nonumber\\
&+&\sum_{k=1}^{t/\epsilon}\delta \bar{z}_{\xi}(k)(\delta z_{\xi}(k)-
\delta z_{\xi}(k-1))\bigg]
-\frac{\epsilon}{N}\sum_{k=1}^{t/\epsilon}
\bigg[
\nonumber\\
&\times & \bigg(1-\frac{\nu}{L}\bigg)
\sum_{\xi,\xi',\xi''}Q_{\xi',\xi} r(\xi)[\delta\bar{z}_{\xi'}(k)-
\delta\bar{z}_{\xi''}(k)]\nonumber\\
&\times &[\delta\bar{z}_{\xi}(k)N^{2}P_{\xi}P_{\xi''}\nonumber\\
&+& N P_{\xi}\delta z_{\xi''}(k-1) + N P_{\xi''}\delta z_{\xi}(k-1)]
\nonumber\\
&+&\sum_{\xi,\xi'}d(\xi')[\delta\bar{z}_{\xi}(k)-\delta\bar{z}_{\xi'}(k)]
[\delta\bar{z}_{\xi}(k) N^{2}P_{\xi}P_{\xi'}\nonumber\\
&+& N P_{\xi}\delta z_{\xi'}(k-1) + N P_{\xi'}\delta z_{\xi}(k-1)]\bigg]
\nonumber\\
&-&\frac{\nu}{L}\frac{\epsilon}{N}\sum_{k=1}^{t/\epsilon}
\sum_{\xi,\xi^{'},\xi^{''}}\{Q_{\xi^{'},\xi+1}\rho_{+}(L-\xi)
+ Q_{\xi^{'},\xi-1}\nonumber\\
&\times &\rho_{-}\xi \}r(\xi)[\delta\bar{z}_{\xi^{'}}(k)
-\delta\bar{z}_{\xi^{''}}(k)]
[\delta\bar{z}_{\xi}(k)N^2 P_{\xi}P_{\xi^{''}}\nonumber\\
& +& N P_{\xi}
\delta z_{\xi^{''}}(k-1)\nonumber\\
&+& N P_{\xi^{''}}\delta z_{\xi}(k-1)]
+ O[(\delta\bar{z},\delta z)^{3}]
\nonumber\\
&=&\frac{1}{2} X^{T} \Pi^{-1} X + O(X^{3})
\label{eqA19}
\end{eqnarray}
Here, we defined $X^{T} = (\{\delta\bar{z}(0),\delta z(0)\},\ldots,
\{\delta\bar{z}(t/\epsilon),\delta z(t/\epsilon)\})$.
The matrix $\Pi^{-1}$ is tridiagonal by blocks, as in the case of the
parallel model. A similar analysis holds for the Eigen model as well,
with matrices $A$ and $B$ defined as
\begin{eqnarray}
&& L^{-1} [A]_{\xi,\xi^{'}} = \bigg(1-\frac{\nu}{L}\bigg)
\bigg[\sum_{\xi^{''}=0}^{L}
Q_{\xi,\xi^{''}}f(\xi^{''})P_{\xi^{''}}
+ Q_{\xi,\xi^{'}}f(\xi^{'})\nonumber\\
&&-\delta_{\xi,\xi^{'}}\sum_{\xi^{''}=0}^{L}f(\xi^{''})P_{\xi^{''}}
-f(\xi^{'})P_{\xi}\bigg]
+[d(\xi^{'})-d(\xi)]P_{\xi}\nonumber\\
&&+ \delta_{\xi,\xi^{'}}\bigg[
\sum_{\xi_{1}=0}^{L}d(\xi_{1})P_{\xi_{1}} - d(\xi)\bigg]
+\frac{\nu}{L}\bigg[\sum_{\xi^{''}=0}^{L}\bigg(Q_{\xi,\xi^{''}-1}
\rho_{-}\xi^{''}\nonumber\\
&&+ Q_{\xi,\xi^{''}+1}\rho_{+}(L-\xi^{''})\bigg)f(\xi^{''})
P_{\xi^{''}} + \bigg(Q_{\xi,\xi^{'}-1}\rho_{-}
\xi^{'}\nonumber\\
&&+Q_{\xi,\xi^{'}+1}\rho_{+}(L-\xi^{'})  \bigg)f(\xi^{'})
-\delta_{\xi,\xi^{'}}\sum_{\xi^{''}=0}^{L}\bigg(\rho_{+}(L-\xi^{''})
\nonumber\\
&&+\rho_{-}\xi^{''}  \bigg)f(\xi^{''})P_{\xi^{''}}
-\bigg(\rho_{+}(L-\xi^{'}) + \rho_{-}\xi^{'}  \bigg)f(\xi^{'})P_{\xi}
\bigg]
\label{eqA20}
\end{eqnarray}

\begin{eqnarray}
&&L^{-1} [B]_{\xi,\xi^{'}} =N\bigg\{\bigg(1-\frac{\nu}{L} \bigg)
\bigg[Q_{\xi^{'},\xi}
f(\xi) P_{\xi}
+ Q_{\xi,\xi^{'}} f(\xi^{'}) P_{\xi^{'}}\nonumber\\
&&-(f(\xi)+ f(\xi^{'}))P_{\xi}P_{\xi^{'}} \bigg]
+2\bigg(\sum_{\xi_{1}=0}^{L}d(\xi_{1})P_{\xi_{1}}\bigg)
P_{\xi}
\delta_{\xi,\xi^{'}}
\nonumber\\
&&
+\frac{\nu}{L}\bigg[\bigg(Q_{\xi^{'},\xi+1}\rho_{+}(L-\xi)
+ Q_{\xi^{'},\xi-1}\rho_{-}\xi \bigg)
f(\xi) P_{\xi}
\nonumber\\
&&
+\bigg(Q_{\xi,\xi^{'}+1}\rho_{+}(L-\xi^{'})
+ Q_{\xi,\xi^{'}-1}\rho_{-}
\xi^{'} \bigg)f(\xi^{'}) P_{\xi^{'}}\nonumber\\
&&-\bigg[\bigg(\rho_{+}(L-\xi) + \rho_{-}\xi \bigg)f(\xi)
+\bigg(\rho_{+}(L-\xi^{'})
+\rho_{-}\xi^{'}\bigg)\nonumber\\
&&\times f(\xi^{'})\bigg]P_{\xi}P_{\xi^{'}}
-(d(\xi)
+ d(\xi^{'}))P_{\xi} P_{\xi^{'}}
\bigg\}
\label{eqA21}
\end{eqnarray}

A recursion relation identical to Eq.\ (\ref{eqA15})
is obtained, which in the
continuous time limit $\epsilon \rightarrow 0$ 
yields a Lyapunov equation for the matrix $C$,
\begin{eqnarray}
\frac{d}{dt}C = B + A C + C A^{T}
\label{eqA22}
\end{eqnarray}
with initial condition $C_{\xi,\xi^{'}}=-\delta_{\xi,\xi^{'}}n_{\xi}^{0}$.

\bibliography{finite_population_long}

\begin{thebibliography}{42}
\expandafter\ifx\csname natexlab\endcsname\relax\def\natexlab#1{#1}\fi
\expandafter\ifx\csname bibnamefont\endcsname\relax
  \def\bibnamefont#1{#1}\fi
\expandafter\ifx\csname bibfnamefont\endcsname\relax
  \def\bibfnamefont#1{#1}\fi
\expandafter\ifx\csname citenamefont\endcsname\relax
  \def\citenamefont#1{#1}\fi
\expandafter\ifx\csname url\endcsname\relax
  \def\url#1{\texttt{#1}}\fi
\expandafter\ifx\csname urlprefix\endcsname\relax\def\urlprefix{URL }\fi
\providecommand{\bibinfo}[2]{#2}
\providecommand{\eprint}[2][]{\url{#2}}

\bibitem[{\citenamefont{Moya et~al.}(2000)\citenamefont{Moya, Elena, Bracho,
  Miralles, and Barrio}}]{Moya00}
\bibinfo{author}{\bibfnamefont{A.}~\bibnamefont{Moya}},
  \bibinfo{author}{\bibfnamefont{S.~F.} \bibnamefont{Elena}},
  \bibinfo{author}{\bibfnamefont{A.}~\bibnamefont{Bracho}},
  \bibinfo{author}{\bibfnamefont{R.}~\bibnamefont{Miralles}}, \bibnamefont{and}
  \bibinfo{author}{\bibfnamefont{E.}~\bibnamefont{Barrio}},
  \bibinfo{journal}{Proc. Natl. Acad. Sci. USA} \textbf{\bibinfo{volume}{97}},
  \bibinfo{pages}{6967} (\bibinfo{year}{2000}).

\bibitem[{\citenamefont{Zhang}(1997)}]{Zhang97}
\bibinfo{author}{\bibfnamefont{Y.-C.} \bibnamefont{Zhang}},
  \bibinfo{journal}{Phys. Rev. E} \textbf{\bibinfo{volume}{55}},
  \bibinfo{pages}{R3817} (\bibinfo{year}{1997}).

\bibitem[{\citenamefont{Alves and Fontanari}(1998)}]{Alves98}
\bibinfo{author}{\bibfnamefont{D.}~\bibnamefont{Alves}} \bibnamefont{and}
  \bibinfo{author}{\bibfnamefont{J.~F.} \bibnamefont{Fontanari}},
  \bibinfo{journal}{Phys. Rev. E} \textbf{\bibinfo{volume}{57}},
  \bibinfo{pages}{7008} (\bibinfo{year}{1998}).

\bibitem[{\citenamefont{Muller}(1964)}]{Muller64}
\bibinfo{author}{\bibfnamefont{H.~J.} \bibnamefont{Muller}},
  \bibinfo{journal}{Mutation Research} \textbf{\bibinfo{volume}{1}},
  \bibinfo{pages}{2} (\bibinfo{year}{1964}).

\bibitem[{\citenamefont{Arjan et~al.}(2007)\citenamefont{Arjan, de~Visser, and
  Elena}}]{Arjan07}
\bibinfo{author}{\bibfnamefont{J.}~\bibnamefont{Arjan}},
  \bibinfo{author}{\bibfnamefont{G.~M.} \bibnamefont{de~Visser}},
  \bibnamefont{and} \bibinfo{author}{\bibfnamefont{S.~F.} \bibnamefont{Elena}},
  \bibinfo{journal}{Nature {R}ev. {G}enet.} \textbf{\bibinfo{volume}{8}},
  \bibinfo{pages}{139} (\bibinfo{year}{2007}).

\bibitem[{\citenamefont{Kondrashov}(1988)}]{Kondrashov88}
\bibinfo{author}{\bibfnamefont{A.~S.} \bibnamefont{Kondrashov}},
  \bibinfo{journal}{Nature} \textbf{\bibinfo{volume}{336}},
  \bibinfo{pages}{435} (\bibinfo{year}{1988}).

\bibitem[{\citenamefont{Chao}(1992)}]{Chao92}
\bibinfo{author}{\bibfnamefont{L.}~\bibnamefont{Chao}},
  \bibinfo{journal}{Trends. {E}col. {E}vol.} \textbf{\bibinfo{volume}{7}},
  \bibinfo{pages}{147} (\bibinfo{year}{1992}).

\bibitem[{\citenamefont{Chao et~al.}(1997)\citenamefont{Chao, Tran, and
  Tran}}]{Chao97}
\bibinfo{author}{\bibfnamefont{L.}~\bibnamefont{Chao}},
  \bibinfo{author}{\bibfnamefont{T.~T.} \bibnamefont{Tran}}, \bibnamefont{and}
  \bibinfo{author}{\bibfnamefont{T.~T.} \bibnamefont{Tran}},
  \bibinfo{journal}{Genetics} \textbf{\bibinfo{volume}{147}},
  \bibinfo{pages}{983} (\bibinfo{year}{1997}).

\bibitem[{\citenamefont{Rice and Chippindale}(2001)}]{Rice01}
\bibinfo{author}{\bibfnamefont{W.~R.} \bibnamefont{Rice}} \bibnamefont{and}
  \bibinfo{author}{\bibfnamefont{A.~K.} \bibnamefont{Chippindale}},
  \bibinfo{journal}{Science} \textbf{\bibinfo{volume}{294}},
  \bibinfo{pages}{555} (\bibinfo{year}{2001}).

\bibitem[{\citenamefont{Adami}(2006)}]{Adami06}
\bibinfo{author}{\bibfnamefont{C.}~\bibnamefont{Adami}},
  \bibinfo{journal}{{N}at. {R}ev. {G}enet.} \textbf{\bibinfo{volume}{7}},
  \bibinfo{pages}{109} (\bibinfo{year}{2006}).

\bibitem[{\citenamefont{Otto and Lenormand}(2002)}]{Otto02}
\bibinfo{author}{\bibfnamefont{S.~P.} \bibnamefont{Otto}} \bibnamefont{and}
  \bibinfo{author}{\bibfnamefont{T.}~\bibnamefont{Lenormand}},
  \bibinfo{journal}{Nature Rev. Genet.} \textbf{\bibinfo{volume}{3}},
  \bibinfo{pages}{252} (\bibinfo{year}{2002}).

\bibitem[{\citenamefont{Kondrashov}(1982)}]{Kondrashov82}
\bibinfo{author}{\bibfnamefont{A.~S.} \bibnamefont{Kondrashov}},
  \bibinfo{journal}{{G}enet. {R}es.} \textbf{\bibinfo{volume}{42}},
  \bibinfo{pages}{325} (\bibinfo{year}{1982}).

\bibitem[{\citenamefont{Kouyos et~al.}(2006)\citenamefont{Kouyos, Otto, and
  Bonhoeffer}}]{Kouyos06}
\bibinfo{author}{\bibfnamefont{R.~D.} \bibnamefont{Kouyos}},
  \bibinfo{author}{\bibfnamefont{S.~P.} \bibnamefont{Otto}}, \bibnamefont{and}
  \bibinfo{author}{\bibfnamefont{S.}~\bibnamefont{Bonhoeffer}},
  \bibinfo{journal}{Genetics} \textbf{\bibinfo{volume}{173}},
  \bibinfo{pages}{589} (\bibinfo{year}{2006}).

\bibitem[{\citenamefont{Kouyos et~al.}(2007)\citenamefont{Kouyos, Silander, and
  Bonhoeffer}}]{Kouyos07}
\bibinfo{author}{\bibfnamefont{R.~D.} \bibnamefont{Kouyos}},
  \bibinfo{author}{\bibfnamefont{O.~K.} \bibnamefont{Silander}},
  \bibnamefont{and}
  \bibinfo{author}{\bibfnamefont{S.}~\bibnamefont{Bonhoeffer}},
  \bibinfo{journal}{Trends {E}col. {E}vol.} \textbf{\bibinfo{volume}{22}},
  \bibinfo{pages}{310} (\bibinfo{year}{2007}).

\bibitem[{\citenamefont{Rouzine et~al.}(2003)\citenamefont{Rouzine, Wakeley,
  and Coffin}}]{Rouzine03}
\bibinfo{author}{\bibfnamefont{I.~M.} \bibnamefont{Rouzine}},
  \bibinfo{author}{\bibfnamefont{J.}~\bibnamefont{Wakeley}}, \bibnamefont{and}
  \bibinfo{author}{\bibfnamefont{J.~M.} \bibnamefont{Coffin}},
  \bibinfo{journal}{Proc. Natl. Acad. Sci. USA} \textbf{\bibinfo{volume}{100}},
  \bibinfo{pages}{587} (\bibinfo{year}{2003}).

\bibitem[{\citenamefont{Rouzine et~al.}(2008)\citenamefont{Rouzine, Brunet, and
  Wilke}}]{Rouzine08}
\bibinfo{author}{\bibfnamefont{I.~M.} \bibnamefont{Rouzine}},
  \bibinfo{author}{\bibfnamefont{E.}~\bibnamefont{Brunet}}, \bibnamefont{and}
  \bibinfo{author}{\bibfnamefont{C.~O.} \bibnamefont{Wilke}},
  \bibinfo{journal}{Theor. Popul. Biol.} \textbf{\bibinfo{volume}{73}},
  \bibinfo{pages}{24} (\bibinfo{year}{2008}).

\bibitem[{\citenamefont{Crow and Kimura}(1970)}]{Kimura70}
\bibinfo{author}{\bibfnamefont{J.~F.} \bibnamefont{Crow}} \bibnamefont{and}
  \bibinfo{author}{\bibfnamefont{M.}~\bibnamefont{Kimura}},
  \emph{\bibinfo{title}{An introduction to population genetics theory}}
  (\bibinfo{publisher}{Harper and Row}, \bibinfo{address}{New York},
  \bibinfo{year}{1970}).

\bibitem[{\citenamefont{Eigen and Schuster}(1971)}]{Eigen71}
\bibinfo{author}{\bibfnamefont{M.}~\bibnamefont{Eigen}} \bibnamefont{and}
  \bibinfo{author}{\bibfnamefont{P.}~\bibnamefont{Schuster}},
  \bibinfo{journal}{Naturwissenschaften} \textbf{\bibinfo{volume}{58}},
  \bibinfo{pages}{465} (\bibinfo{year}{1971}).

\bibitem[{\citenamefont{Eigen et~al.}(1988)\citenamefont{Eigen, McCaskill, and
  Schuster}}]{Eigen88}
\bibinfo{author}{\bibfnamefont{M.}~\bibnamefont{Eigen}},
  \bibinfo{author}{\bibfnamefont{J.}~\bibnamefont{McCaskill}},
  \bibnamefont{and} \bibinfo{author}{\bibfnamefont{P.}~\bibnamefont{Schuster}},
  \bibinfo{journal}{J. Phys. Chem.} \textbf{\bibinfo{volume}{92}},
  \bibinfo{pages}{6881} (\bibinfo{year}{1988}).

\bibitem[{\citenamefont{Eigen et~al.}(1989)\citenamefont{Eigen, McCaskill, and
  Schuster}}]{Eigen89}
\bibinfo{author}{\bibfnamefont{M.}~\bibnamefont{Eigen}},
  \bibinfo{author}{\bibfnamefont{J.}~\bibnamefont{McCaskill}},
  \bibnamefont{and} \bibinfo{author}{\bibfnamefont{P.}~\bibnamefont{Schuster}},
  \bibinfo{journal}{Adv. Chem. Phys.} \textbf{\bibinfo{volume}{75}},
  \bibinfo{pages}{149} (\bibinfo{year}{1989}).

\bibitem[{\citenamefont{Biebricher and Eigen}(2005)}]{Biebricher05}
\bibinfo{author}{\bibfnamefont{C.~K.} \bibnamefont{Biebricher}}
  \bibnamefont{and} \bibinfo{author}{\bibfnamefont{M.}~\bibnamefont{Eigen}},
  \bibinfo{journal}{Virus Res.} \textbf{\bibinfo{volume}{107}},
  \bibinfo{pages}{117} (\bibinfo{year}{2005}).

\bibitem[{\citenamefont{Tarazona}(1992)}]{Tarazona92}
\bibinfo{author}{\bibfnamefont{P.}~\bibnamefont{Tarazona}},
  \bibinfo{journal}{Phys. Rev. A} \textbf{\bibinfo{volume}{45}},
  \bibinfo{pages}{6038} (\bibinfo{year}{1992}).

\bibitem[{\citenamefont{Leuthausser}(1987)}]{Leuthausser87}
\bibinfo{author}{\bibfnamefont{I.}~\bibnamefont{Leuthausser}},
  \bibinfo{journal}{J. Stat. Phys.} \textbf{\bibinfo{volume}{48}},
  \bibinfo{pages}{343} (\bibinfo{year}{1987}).

\bibitem[{\citenamefont{Franz and Peliti}(1997)}]{Franz97}
\bibinfo{author}{\bibfnamefont{S.}~\bibnamefont{Franz}} \bibnamefont{and}
  \bibinfo{author}{\bibfnamefont{L.}~\bibnamefont{Peliti}},
  \bibinfo{journal}{J. Phys. A: Math. Gen.} \textbf{\bibinfo{volume}{30}},
  \bibinfo{pages}{4481} (\bibinfo{year}{1997}).

\bibitem[{\citenamefont{Park and Deem}(2006)}]{Park06}
\bibinfo{author}{\bibfnamefont{J.-M.} \bibnamefont{Park}} \bibnamefont{and}
  \bibinfo{author}{\bibfnamefont{M.~W.} \bibnamefont{Deem}},
  \bibinfo{journal}{J. Stat. Phys.} \textbf{\bibinfo{volume}{123}},
  \bibinfo{pages}{975} (\bibinfo{year}{2006}).

\bibitem[{\citenamefont{Saakian et~al.}(2006)\citenamefont{Saakian,
  {Mu\~{n}oz}, Hu, and Deem}}]{Saakian06}
\bibinfo{author}{\bibfnamefont{D.~B.} \bibnamefont{Saakian}},
  \bibinfo{author}{\bibfnamefont{E.}~\bibnamefont{{Mu\~{n}oz}}},
  \bibinfo{author}{\bibfnamefont{C.-K.} \bibnamefont{Hu}}, \bibnamefont{and}
  \bibinfo{author}{\bibfnamefont{M.~W.} \bibnamefont{Deem}},
  \bibinfo{journal}{Phys. Rev. E} \textbf{\bibinfo{volume}{73}},
  \bibinfo{pages}{041913} (\bibinfo{year}{2006}).

\bibitem[{\citenamefont{Mu{\~n}oz et~al.}(2008)\citenamefont{Mu{\~n}oz, Park,
  and Deem}}]{Munoz08}
\bibinfo{author}{\bibfnamefont{E.}~\bibnamefont{Mu{\~n}oz}},
  \bibinfo{author}{\bibfnamefont{J.-M.} \bibnamefont{Park}}, \bibnamefont{and}
  \bibinfo{author}{\bibfnamefont{M.~W.} \bibnamefont{Deem}},
  \bibinfo{journal}{Phys. Rev. E} \textbf{\bibinfo{volume}{78}},
  \bibinfo{pages}{061921} (\bibinfo{year}{2008}).

\bibitem[{\citenamefont{Mu{\~n}oz et~al.}(2009)\citenamefont{Mu{\~n}oz, Park,
  and Deem}}]{Munoz09}
\bibinfo{author}{\bibfnamefont{E.}~\bibnamefont{Mu{\~n}oz}},
  \bibinfo{author}{\bibfnamefont{J.-M.} \bibnamefont{Park}}, \bibnamefont{and}
  \bibinfo{author}{\bibfnamefont{M.~W.} \bibnamefont{Deem}},
  \bibinfo{journal}{J. Stat. Phys.} \textbf{\bibinfo{volume}{135}},
  \bibinfo{pages}{429} (\bibinfo{year}{2009}).

\bibitem[{\citenamefont{Domingo et~al.}(1978)\citenamefont{Domingo, Sabo,
  Taniguchi, and Weissman}}]{Domingo78}
\bibinfo{author}{\bibfnamefont{E.}~\bibnamefont{Domingo}},
  \bibinfo{author}{\bibfnamefont{D.}~\bibnamefont{Sabo}},
  \bibinfo{author}{\bibfnamefont{T.}~\bibnamefont{Taniguchi}},
  \bibnamefont{and} \bibinfo{author}{\bibfnamefont{C.}~\bibnamefont{Weissman}},
  \bibinfo{journal}{Cell} \textbf{\bibinfo{volume}{13}}, \bibinfo{pages}{735}
  (\bibinfo{year}{1978}).

\bibitem[{\citenamefont{Domingo et~al.}(2005)\citenamefont{Domingo, Escarmis,
  Lazaro, and Manrubia}}]{Domingo05}
\bibinfo{author}{\bibfnamefont{E.}~\bibnamefont{Domingo}},
  \bibinfo{author}{\bibfnamefont{C.}~\bibnamefont{Escarmis}},
  \bibinfo{author}{\bibfnamefont{E.}~\bibnamefont{Lazaro}}, \bibnamefont{and}
  \bibinfo{author}{\bibfnamefont{S.~C.} \bibnamefont{Manrubia}},
  \bibinfo{journal}{Virus Res.} \textbf{\bibinfo{volume}{107}},
  \bibinfo{pages}{129} (\bibinfo{year}{2005}).

\bibitem[{\citenamefont{Ortin et~al.}(1980)\citenamefont{Ortin, Najera, Lopez,
  Davila, and Domingo}}]{Ortin80}
\bibinfo{author}{\bibfnamefont{J.}~\bibnamefont{Ortin}},
  \bibinfo{author}{\bibfnamefont{R.}~\bibnamefont{Najera}},
  \bibinfo{author}{\bibfnamefont{C.}~\bibnamefont{Lopez}},
  \bibinfo{author}{\bibfnamefont{M.}~\bibnamefont{Davila}}, \bibnamefont{and}
  \bibinfo{author}{\bibfnamefont{E.}~\bibnamefont{Domingo}},
  \bibinfo{journal}{Gene} \textbf{\bibinfo{volume}{11}}, \bibinfo{pages}{319}
  (\bibinfo{year}{1980}).

\bibitem[{\citenamefont{Domingo et~al.}(1985)\citenamefont{Domingo,
  Martinez-Salas, Sobrino, de~la Torre, Portela, Ortin, Lopez-Galindez,
  Perez-Bre{\~n}a, Villanueva, Najera et~al.}}]{Domingo85}
\bibinfo{author}{\bibfnamefont{E.}~\bibnamefont{Domingo}},
  \bibinfo{author}{\bibfnamefont{E.}~\bibnamefont{Martinez-Salas}},
  \bibinfo{author}{\bibfnamefont{F.}~\bibnamefont{Sobrino}},
  \bibinfo{author}{\bibfnamefont{J.~C.} \bibnamefont{de~la Torre}},
  \bibinfo{author}{\bibfnamefont{A.}~\bibnamefont{Portela}},
  \bibinfo{author}{\bibfnamefont{J.}~\bibnamefont{Ortin}},
  \bibinfo{author}{\bibfnamefont{C.}~\bibnamefont{Lopez-Galindez}},
  \bibinfo{author}{\bibfnamefont{P.}~\bibnamefont{Perez-Bre{\~n}a}},
  \bibinfo{author}{\bibfnamefont{N.}~\bibnamefont{Villanueva}},
  \bibinfo{author}{\bibfnamefont{R.}~\bibnamefont{Najera}},
  \bibnamefont{et~al.}, \bibinfo{journal}{Gene} \textbf{\bibinfo{volume}{40}},
  \bibinfo{pages}{1} (\bibinfo{year}{1985}).

\bibitem[{\citenamefont{Cohen et~al.}(2005)\citenamefont{Cohen, Kessler, and
  Levine}}]{Cohen05}
\bibinfo{author}{\bibfnamefont{E.}~\bibnamefont{Cohen}},
  \bibinfo{author}{\bibfnamefont{D.~A.} \bibnamefont{Kessler}},
  \bibnamefont{and} \bibinfo{author}{\bibfnamefont{H.}~\bibnamefont{Levine}},
  \bibinfo{journal}{Phys. Rev. Lett.} \textbf{\bibinfo{volume}{94}},
  \bibinfo{pages}{098102} (\bibinfo{year}{2005}).

\bibitem[{\citenamefont{Park and Deem}(2007)}]{Park07}
\bibinfo{author}{\bibfnamefont{J.-M.} \bibnamefont{Park}} \bibnamefont{and}
  \bibinfo{author}{\bibfnamefont{M.~W.} \bibnamefont{Deem}},
  \bibinfo{journal}{Phys. Rev. Lett.} \textbf{\bibinfo{volume}{98}},
  \bibinfo{pages}{058101} (\bibinfo{year}{2007}).

\bibitem[{\citenamefont{Peliti}(1985)}]{Peliti85}
\bibinfo{author}{\bibfnamefont{L.}~\bibnamefont{Peliti}}, \bibinfo{journal}{J.
  Physique} \textbf{\bibinfo{volume}{46}}, \bibinfo{pages}{1469}
  (\bibinfo{year}{1985}).

\bibitem[{\citenamefont{Mattis and Glasser}(1998)}]{Mattis98}
\bibinfo{author}{\bibfnamefont{D.~C.} \bibnamefont{Mattis}} \bibnamefont{and}
  \bibinfo{author}{\bibfnamefont{M.~L.} \bibnamefont{Glasser}},
  \bibinfo{journal}{Rev. Mod. Phys.} \textbf{\bibinfo{volume}{70}},
  \bibinfo{pages}{979} (\bibinfo{year}{1998}).

\bibitem[{\citenamefont{Saakian}(2007)}]{Saakian07}
\bibinfo{author}{\bibfnamefont{D.~B.} \bibnamefont{Saakian}}
  (\bibinfo{year}{2007}), \bibinfo{note}{personal communication}.

\bibitem[{\citenamefont{Gillespie}(1976)}]{Gillespie76}
\bibinfo{author}{\bibfnamefont{D.~T.} \bibnamefont{Gillespie}},
  \bibinfo{journal}{J. Comput. Phys.} \textbf{\bibinfo{volume}{22}},
  \bibinfo{pages}{403} (\bibinfo{year}{1976}).

\bibitem[{\citenamefont{Bortz et~al.}(1995)\citenamefont{Bortz, Kalos, and
  Lebowitz}}]{Lebowitz75}
\bibinfo{author}{\bibfnamefont{A.~B.} \bibnamefont{Bortz}},
  \bibinfo{author}{\bibfnamefont{M.~H.} \bibnamefont{Kalos}}, \bibnamefont{and}
  \bibinfo{author}{\bibfnamefont{J.~L.} \bibnamefont{Lebowitz}},
  \bibinfo{journal}{J. Comput. Phys.} \textbf{\bibinfo{volume}{17}},
  \bibinfo{pages}{10} (\bibinfo{year}{1995}).

\bibitem[{\citenamefont{Bogarad and Deem}(1999)}]{Bogarad99}
\bibinfo{author}{\bibfnamefont{L.~D.} \bibnamefont{Bogarad}} \bibnamefont{and}
  \bibinfo{author}{\bibfnamefont{M.~W.} \bibnamefont{Deem}},
  \bibinfo{journal}{Proc. Natl. Acad. Sci. USA} \textbf{\bibinfo{volume}{96}},
  \bibinfo{pages}{2591} (\bibinfo{year}{1999}).

\bibitem[{\citenamefont{Earl and Deem}(2004)}]{Earl04}
\bibinfo{author}{\bibfnamefont{D.~J.} \bibnamefont{Earl}} \bibnamefont{and}
  \bibinfo{author}{\bibfnamefont{M.~W.} \bibnamefont{Deem}},
  \bibinfo{journal}{Proc. Natl. Acad. Sci. USA} \textbf{\bibinfo{volume}{101}},
  \bibinfo{pages}{11531} (\bibinfo{year}{2004}).

\bibitem[{\citenamefont{Sun and Deem}(2007)}]{Sun07}
\bibinfo{author}{\bibfnamefont{J.}~\bibnamefont{Sun}} \bibnamefont{and}
  \bibinfo{author}{\bibfnamefont{M.~W.} \bibnamefont{Deem}},
  \bibinfo{journal}{Phys. Rev. Lett.} \textbf{\bibinfo{volume}{99}},
  \bibinfo{pages}{228107} (\bibinfo{year}{2007}).

\end{thebibliography}

\end{document}